\documentclass[]{article}

\usepackage[utf8]{inputenc}
\usepackage{booktabs}
\usepackage{graphicx}
\usepackage{array}
\usepackage[showframe=false]{geometry}
\usepackage{changepage}
\usepackage[]{graphicx}
\usepackage{caption}
\usepackage{subcaption}
\usepackage[toc,page]{appendix}
\usepackage{bm}
\usepackage{float}
\usepackage{amsmath}
\usepackage{mathtools}   
\usepackage{authblk}
\usepackage[colorlinks=true, allcolors=blue]{hyperref}
\usepackage{longtable}
\numberwithin{equation}{section}
\usepackage[dvipsnames]{xcolor}
\usepackage{afterpage}
\usepackage{makecell}
\usepackage{amssymb}
\usepackage{hyperref}
\usepackage{cite}
\usepackage{cleveref}
\usepackage{lmodern}
\usepackage{hyperref}
\usepackage{titlesec}
\usepackage{xcolor}
\usepackage[normalem]{ulem}

\titleformat{\subsection}[hang]{\normalfont\normalsize}{\normalfont\normalsize\thesubsection}{1em}{}

\newcommand{\refig}[1]{Fig.~\ref{#1}}
\newcommand{\ret}[1]{Table~\ref{#1}}
\crefname{equation}{Eq.}{Eqs.}
\crefrangelabelformat{equation}{(#3#1#4)--(#5#2#6)}

\usepackage{fancyhdr}
\fancypagestyle{preprint}{%
	\fancyhf{}
	\fancyhead[R]{\small MITP-26-063}
	\cfoot{\thepage}
	
}

\title{\textbf{The ratio of gradient-flow scales $\sqrt{t_0}/w_0$ in $2+1$ flavor lattice QCD}}
\author[1,2,3]{Simon Kuberski}
\affil[1]{PRISMA$^{++}$ Cluster of Excellence and Institut f\"ur Kernphysik,
Johannes Gutenberg-Universit\"at Mainz, 55099 Mainz, Germany}
\affil[2]{Helmholtz-Institut Mainz, Johannes Gutenberg-Universit\"at Mainz, 55099 Mainz, Germany}
\affil[3]{GSI Helmholtzzentrum f\"ur Schwerionenforschung, 64291 Darmstadt, Germany}
\author[4]{Alejandro S\'aez}
\affil[4]{Instituto de Física Corpuscular (IFIC), CSIC-Universitat de València, 46071, Valencia, Spain}

\begin{document}
\maketitle
\thispagestyle{preprint}

\begin{abstract}

We determine the ratio $\sqrt{t_0}/w_0$ of gradient-flow scales in isospin-symmetric $2+1$-flavor QCD at the physical point and in the continuum limit. 
Our analysis uses CLS gauge ensembles at five lattice spacings and combines Wilson and Zeuthen flow data to constrain the continuum extrapolation at the flavor-SU(3)-symmetric point. 
We obtain $\sqrt{t_0}/w_0=0.8264(26)$ in the Edinburgh consensus scheme, corresponding to a precision of about $0.31\%$.
This result relates determinations quoted in terms of either flow scale and helps assess tensions among published results.

\end{abstract}

\section{Introduction}
Lattice QCD provides precision results for observables that are highly relevant to tests of the Standard Model and to particle physics phenomenology.
Prime examples are recent determinations of the strong coupling constant~\cite{DallaBrida:2025ymw}, the hadronic vacuum-polarization contribution to the anomalous magnetic moment $a_\mu^{\rm HVP}$~\cite{Aliberti:2025beg}, and decay constants relevant to leptonic meson decays~\cite{FlavourLatticeAveragingGroupFLAG:2024oxs}.

To relate observables computed in lattice units to physical units, the lattice spacing has to be set by requiring an observable to take its physical value.
The uncertainty from this conversion then enters all dimensionful observables computed on the lattice, but prominently also some dimensionless quantities such as $a_\mu^{\rm HVP}$~\cite{DellaMorte:2017dyu}.

The nature of this uncertainty depends on the observable chosen to set the scale.
Whereas hadron masses can be obtained reliably in experiment, their extraction on the lattice can be subject to substantial systematic uncertainties from excited-state contamination and rapidly increasing statistical errors. 
Decay constants of light pseudoscalar mesons, in turn, are precision quantities on the lattice, but relating them to experiment in full QCD+QED requires a prescription.

Intermediate scale setting quantities derived from the gradient flow, such as $t_0$~\cite{Luscher:2010iy} and $w_0$~\cite{BMW:2012hcm} can be obtained precisely and reliably in lattice QCD calculations but have no direct experimental counterparts.
Their physical values must therefore be determined on the lattice through their relation to a physical quantity.

Currently, uncertainties in these determinations make significant or even dominant contributions to the final uncertainty of some precision calculations; see, for example, \cite{DallaBrida:2025ymw,Boccaletti:2024guq,Kuberski:2024pms,Bussone:2023kag}.
In addition, the review by the Flavor Lattice Averaging Group \cite{FlavourLatticeAveragingGroupFLAG:2024oxs} shows tensions between some lattice determinations, which may point to underestimated uncertainties. 
Investigating these tensions is hindered by the fact that some groups determine $t_0$, whereas others report only $w_0$ despite the close relation between the two scales.
At present, only a few results allow the two quantities to be converted into each other \cite{Dowdall:2013rya,ExtendedTwistedMass:2021qui,Larsen:2025wvg}.

While gradient-flow quantities can usually be obtained with high statistical precision, they can have substantial cutoff effects arising from the simulated action and from the definitions of the flow and observable \cite{Ramos:2015baa}. 
This complicates the continuum extrapolation and makes fine lattice spacings important for an accurate result.

In this work, we determine the ratio of flow observables $\sqrt{t_0}/w_0$ in isospin-symmetric QCD with $2+1$ flavors of quarks, based on a set of high-statistics gauge ensembles generated by the CLS initiative.
Our result is obtained with a precision of about 0.31\%, where the uncertainty is dominated by our estimate of systematic effects, mostly from the continuum extrapolation.

Our result is obtained in the scheme defined by the Edinburgh consensus, as recommended in~\cite{FlavourLatticeAveragingGroupFLAG:2024oxs}.
We also provide the derivatives of our result with respect to the observables that define the scheme, such that small variations of the scheme can be applied a posteriori.
We use our result for $\sqrt{t_0}/w_0$ in combination with a recent determination of $\sqrt{t_0}$~\cite{Bussone:2025wlf} on a similar set of gauge configurations to quote a result for $w_0$ in the same scheme.

This paper is structured as follows. Section \ref{sec:setup} introduces the
setup of our calculation and our strategy for approaching the physical point.
Section \ref{sec:continuum} describes the continuum extrapolation of the ratio
at the flavor-SU(3)-symmetric point. In Section \ref{sec:chiral_extrap}, we
perform the remaining chiral interpolation and continuum extrapolation to obtain
the ratio at the physical point. We compare our results with existing
determinations and conclude in Section \ref{sec:conclusions}. In Appendix
\ref{app:tables} we tabulate the observables employed for the analysis as
obtained on the gauge ensembles, while Appendix \ref{app:chisq} discusses
details about the fitting strategy.

\section{Lattice setup}
\label{sec:setup}

\begin{figure}[t!]
  \centering
  \includegraphics[width=.49\textwidth]{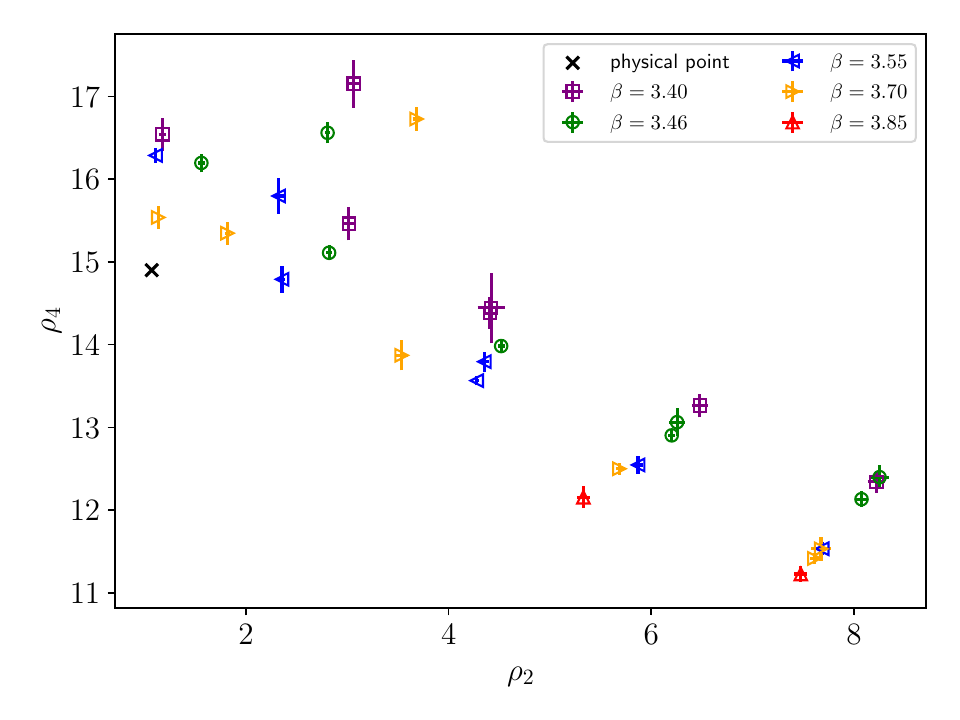}
  \includegraphics[width=.49\textwidth]{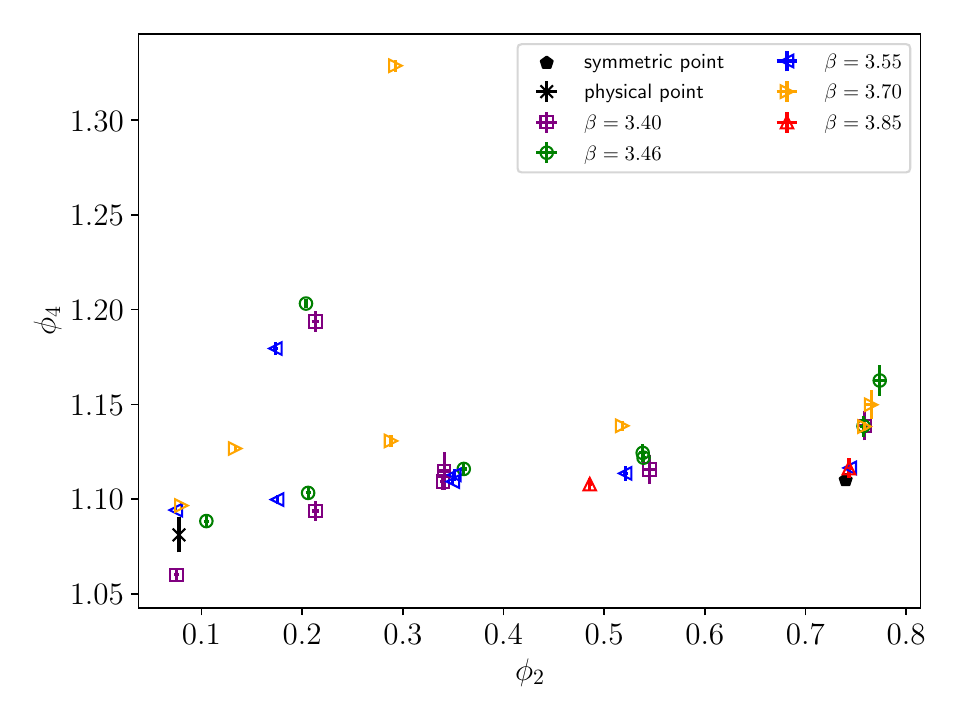}
  \caption{\label{fig:rho_phi}
  \textit{Left panel:} values for $\rho_2$ and $\rho_4$ along the set of CLS ensembles under study.
  The physical point is given with no errors by \cref{eq:edinburgh}, and represented by the black cross.
  \textit{Right panel:} the same, for $\phi_2$ and $\phi_4$, using the Wilson flow values of $t_0$.
  The physical point is given by \cref{eq:edinburgh} and $\sqrt{t_0}=0.1440(7)$ fm \cite{Bussone:2025wlf}, and represented by the black cross, while the symmetric point definition $\phi_2=0.74$, $\phi_4=\frac{3}{2}\phi_2$ (cf. the main text) is represented by a black pentagon.
  The $\rho$- and $\phi$-variables are defined in \cref{eq:rhos,eq:phis}.
  }
\end{figure}

The set of lattice ensembles under study were generated by the Coordinated Lattice Simulation (CLS) initiative \cite{Bruno:2014jqa,Bruno:2016plf,Bali:2016umi,Mohler:2017wnb,Mohler:2020txx,Kuberski:2023zky}.
They employ the tree-level $\mbox{O}(a^2)$-improved Lüscher-Weisz gauge action \cite{Luscher:1984xn,Luscher:1985zq} and non-perturbatively $\mbox{O}(a)$-improved $N_f=2+1$ Wilson fermions \cite{Bulava:2013cta}, with lattice spacings ranging from $a\approx0.085$ fm down to $a\approx0.039$ fm, and pion masses from $m_{\pi}\approx420$ MeV down to their physical value.
Open boundary conditions (OBC) are used in the temporal direction of the gauge fields for most of these ensembles in order to avoid topological freezing as the continuum limit is approached.
We employ ensembles belonging to two different chiral trajectories in the quark-mass plane: one defined by $m_{\rm s}\approx m_{\rm s}^{\rm ph}$; and one defined by a constant value of the bare quark-mass matrix at each value of the bare coupling $\beta$,
\begin{equation}
  \label{eq:chiral_traj_phi4}
{\rm tr}\left(M_q\right)=2m_{\rm ud}+m_{\rm s}={\rm cnst.}\Rightarrow\phi_4\equiv8t_0\left(m_K^2+\frac{1}{2}m_{\pi}^2\right)\approx{\rm cnst}.
\end{equation}
The complete list of ensembles is shown in \ret{tab:ens}.
As kinematical variables we will employ the pion and kaon masses in units of the pion decay constant
\begin{equation}
  \label{eq:rhos}
  \rho_2=\frac{m_{\pi}^2}{f_{\pi}^2},\quad \rho_4=\frac{m_K^2+\frac{1}{2}m_{\pi}^2}{f_{\pi}^2}.
\end{equation}
though it will be convenient to introduce equivalent quantities expressed in units of the flow scale $t_0$,
\begin{equation}
  \label{eq:phis}
  \phi_2=8t_0m_{\pi}^2, \quad   \phi_4=8t_0\left(m_K^2+\frac{1}{2}m_{\pi}^2\right),
\end{equation}
The $\rho$-variables allow to unambiguously follow the Edinburgh Consensus \cite{FlavourLatticeAveragingGroupFLAG:2024oxs} concerning the definition of the physical point in isospin-symmetric QCD (isoQCD),
\begin{equation}
  \label{eq:edinburgh}
m_{\pi}^{\rm ph}=135{\rm\;MeV}, \quad m_K^{\rm ph}=494.6{\rm\;MeV}, \quad f_{\pi}^{\rm ph}=130.5{\rm\;MeV},
\end{equation}
while for the $\phi$-variables the intermediate scale $t_0$ needs to be determined accordingly.
In \refig{fig:rho_phi} we illustrate the values of the $\rho$- and $\phi$-variables for the ensembles under study.
Pion and kaon masses and decay constants are taken from the determination in~\cite{Djukanovic:2024cmq,Beltran:2026ofp} where they have been computed from two-point correlation functions involving the pseudoscalar density and the axial-vector current, as described in~\cite{Ce:2022kxy}. 
The flow scales $t_0$ and $w_0$ have been defined in \cite{Luscher:2010iy,BMW:2012hcm} from the flowed Yang-Mills action density $E(t)$ via
\begin{align}
	\left\{t^2 \langle E(t)\rangle \right\}_{t = t_0} = 0.3\,,
	\qquad 
	\left\{ 
	t \frac{\mathrm{d}}{\mathrm{d}t}\left(t^2 \langle E(t)\rangle \right)
	\right\}_{t = w_0^2} = 0.3\,.
\end{align}
Here, we employ either Wilson flow with the clover discretization of the energy density, or the Zeuthen flow \cite{Ramos:2015baa} with the improved energy density,
\begin{equation}
  \label{eq:E_impr}
E(t)_{\rm impr}=\frac{4}{3}E(t)_{\rm pl}-\frac{1}{3}E(t)_{\rm cl},
\end{equation}
where the definition of the plaquette (pl) and clover (cl) discretization of the energy density can be found e.g.\ in \cite{Ramos:2015baa}.
The improved discretization in \cref{eq:E_impr} has no classical $\mbox{O}(a^2)$ cutoff effects -- those coming from the flow and observable discretizations -- but only receives lattice artefacts coming from the lattice action \cite{Ramos:2015baa} at this order.
We will impose universality, i.e., that both regularizations agree in the continuum limit in order to constrain the continuum extrapolation.
In this study, Zeuthen-flow measurements of $t_0/a^2$ and $w_0/a$ are only available for ensembles at the flavor-SU(3)-symmetric point, defined by $m_{\rm ud}=m_{\rm s}$ leading to $m_K=m_{\pi}\approx420$ MeV, see \ret{tab:ens}.
In order to optimize the use of the available data, we divide the computation of $\sqrt{t_0}/w_0$ at the physical point and continuum limit in two steps.
\begin{itemize}
\item \textit{Continuum limit at the symmetric point:}
We perform a continuum limit extrapolation of $\sqrt{t_0}/w_0|_{\rm sym}$, along the line of constant physics (LCP) defined by $\phi_2=0.74,\;\phi_4=\frac{3}{2}\phi_2=1.11$, which corresponds to $\phi_4^*$ in \cite{Bruno:2016plf,RQCD:2022xux}.
This condition is satisfied by the symmetric point ensembles up to small mass-mistunings that are dealt with by use of linear mass-shifts (cf. \cref{eq:md} below).
We employ both Wilson- and Zeuthen-flow determinations of $\sqrt{t_0}/w_0$ along this LCP, imposing a common continuum limit in order to constrain the extrapolation in a controlled way.
Zeuthen-flow measurements were done using the \verb|LatticeGPU.jl| package \cite{Catumba:2025jae}.
\item \textit{Chiral interpolation to the physical point:}
We perform a chiral interpolation of the ratio $(\sqrt{t_0}/w_0)/(\sqrt{t_0}/w_0|_{\rm sym})$.
In this ratio of ratios, most cutoff effects cancel, leaving residual
mass-dependent effects.
There is therefore a residual continuum limit extrapolation that can be safely dealt with by using Wilson-flow data.
\end{itemize}
Combining the results of both steps, a determination of $\sqrt{t_0}/w_0$ at the physical point and continuum limit is obtained.
We summarize the numerical values of the observables entering this work as obtained on the gauge ensembles in Tables \ref{tab:obs1}--\ref{tab:obs2}, where the masses and decay constants of the pseudoscalar mesons have been computed in \cite{Djukanovic:2024cmq,Beltran:2026ofp}.
Finite-size effects in the pion masses and decay constants are corrected using the resummed finite-volume formulae of Refs.~\cite{Colangelo:2003hf,Colangelo:2005gd}, while kaon masses and decay constants are corrected at NLO in $\chi$PT.

The flow scales correspond to smoothing radii of approximately $0.4$--$0.5$ fm, much smaller than the spatial extents of the ensembles considered here. 
Within chiral perturbation theory, finite-volume corrections to $t_0$ and $w_0$ enter through chiral logarithms, which first appear at NNLO, and are therefore expected to be small \cite{Bar:2013ora}. 
Four pairs of ensembles with matched Lagrangian parameters are available at different volumes. 
Their unshifted Wilson-flow ratios $\sqrt{t_0}/w_0$ show no systematic trend with volume, and the observed differences are not statistically significant. 
We therefore do not resolve finite-volume effects in these quantities at our current precision.
\begin{table}[tbp]
	\centering
	\begin{tabular}{cccccccccc}
		\toprule
		id & $\beta$ & BC & $L/a$ & $T/a$ & $a\,[\mathrm{fm}]$
		& $m_\pi\,[\mathrm{MeV}]$ & $m_K\,[\mathrm{MeV}]$
		& $m_\pi L$ & $L\,[\mathrm{fm}]$ \\
		\midrule
    H101 & $3.40$ & o & $32$ & $96$ & $0.085$ & $427$ & $427$ & 5.8 & $2.7$ \\
    H102 & $3.40$ & o & $32$ & $96$ & $0.085$ & $359$ & $445$ & 4.9 & $2.7$ \\
    H105 & $3.40$ & o & $32$ & $96$ & $0.085$ & $285$ & $470$ & 3.9 & $2.7$ \\
    N101 & $3.40$ & o & $48$ & $128$ & $0.085$ & $281$ & $468$ & 5.8 & $4.1$ \\
    C101 & $3.40$ & o & $48$ & $96$ & $0.085$ & $222$ & $478$ & 4.6 & $4.1$ \\
    C102$^{\dagger}$ & $3.40$ & o & $48$ & $96$ & $0.085$ & $224$ & $505$ & 4.6 & $4.1$ \\
    D150$^{\dagger}$ & $3.40$ & p & $64$ & $128$ & $0.085$ & $131$ & $484$ & 3.6 & $5.4$ \\
    \midrule
    B450 & $3.46$ & p & $32$ & $64$ & $0.075$ & $424$ & $424$ & 5.2 & $2.4$ \\
    H400 & $3.46$ & o & $32$ & $96$ & $0.075$ & $430$ & $430$ & 5.2 & $2.4$ \\
    S400 & $3.46$ & o & $32$ & $128$ & $0.075$ & $356$ & $448$ & 4.3 & $2.4$ \\
    N452 & $3.46$ & p & $48$ & $128$ & $0.075$ & $356$ & $447$ & 6.5 & $3.6$ \\
    N451 & $3.46$ & p & $48$ & $128$ & $0.075$ & $291$ & $469$ & 5.3 & $3.6$ \\
    D450 & $3.46$ & p & $64$ & $128$ & $0.075$ & $220$ & $484$ & 5.3 & $4.8$ \\
    D451$^{\dagger}$ & $3.46$ & p & $64$ & $128$ & $0.075$ & $219$ & $510$ & 5.3 & $4.8$ \\
    D452 & $3.46$ & p & $64$ & $128$ & $0.075$ & $156$ & $491$ & 3.8 & $4.8$ \\
    \midrule
    N202 & $3.55$ & o & $48$ & $128$ & $0.063$ & $420$ & $420$ & 6.4 & $3.0$ \\
    N203 & $3.55$ & o & $48$ & $128$ & $0.063$ & $352$ & $451$ & 5.4 & $3.0$ \\
    N200 & $3.55$ & o & $48$ & $128$ & $0.063$ & $289$ & $472$ & 4.4 & $3.0$ \\
    D251 & $3.55$ & p & $64$ & $128$ & $0.063$ & $288$ & $471$ & 4.2 & $4.0$ \\
    D200 & $3.55$ & o & $64$ & $128$ & $0.063$ & $205$ & $490$ & 5.9 & $4.0$ \\
    D201$^{\dagger}$ & $3.55$ & o & $64$ & $128$ & $0.063$ & $202$ & $507$ & 4.2 & $4.1$ \\
    E250$^{\dagger}$ & $3.55$ & p & $96$ & $192$ & $0.063$ & $133$ & $499$ & 4.1 & $6.0$ \\
    \midrule
    J307 & $3.70$ & o & $64$ & $192$ & $0.049$ & $423$ & $423$ & 6.7 & $3.1$ \\
    N300 & $3.70$ & o & $48$ & $128$ & $0.049$ & $428$ & $428$ & 5.1 & $2.4$ \\
    J306 & $3.70$ & o & $64$ & $192$ & $0.049$ & $350$ & $456$ & 5.6 & $3.1$ \\
    J303 & $3.70$ & o & $64$ & $192$ & $0.049$ & $260$ & $483$ & 4.1 & $3.1$ \\
    J304$^{\dagger}$ & $3.70$ & o & $64$ & $192$ & $0.049$ & $264$ & $531$ & 4.2 & $3.1$ \\
    E300 & $3.70$ & o & $96$ & $192$ & $0.049$ & $177$ & $499$ & 4.2 & $4.7$ \\
    F300$^{\dagger}$ & $3.70$ & o & $128$ & $256$ & $0.049$ & $137$ & $498$ & 4.4 & $6.3$ \\
    \midrule
    J500 & $3.85$ & o & $64$ & $192$ & $0.038$ & $424$ & $424$ & 5.2 & $2.4$ \\
    J501 & $3.85$ & o & $64$ & $192$ & $0.038$ & $343$ & $457$ & 4.2 & $2.4$ \\
    \bottomrule
    \end{tabular}
    \caption{\label{tab:ens}
    List of CLS ensembles employed in this work.
    Columns show the following parameters: the inverse bare gauge coupling $\beta$; whether open (o) or periodic (p) boundary conditions are used for the temporal direction of the gauge fields; the spatial and temporal lattice extents; the lattice spacing, based on \cite{Bussone:2025wlf}; approximate values of the pion and kaon masses in physical units; and the physical size of the lattice.
    Ensembles marked with a dagger belong to the chiral trajectory where $m_{\rm s}\approx m_{\rm s}^{\rm ph}$, while the rest belong to the ${\rm tr}\left(M_q\right)={\rm cnst.}$ trajectory.
    }
\end{table}

\section{Continuum limit at the symmetric point}
\label{sec:continuum}

In this section we present the extrapolation of the ratio $\sqrt{t_0}/w_0$ to the continuum limit along the SU(3)-symmetric point.
\subsection{\textit{Symmetric point tuning}}
\begin{figure}[t!]
  \centering
  \includegraphics[width=.49\textwidth]{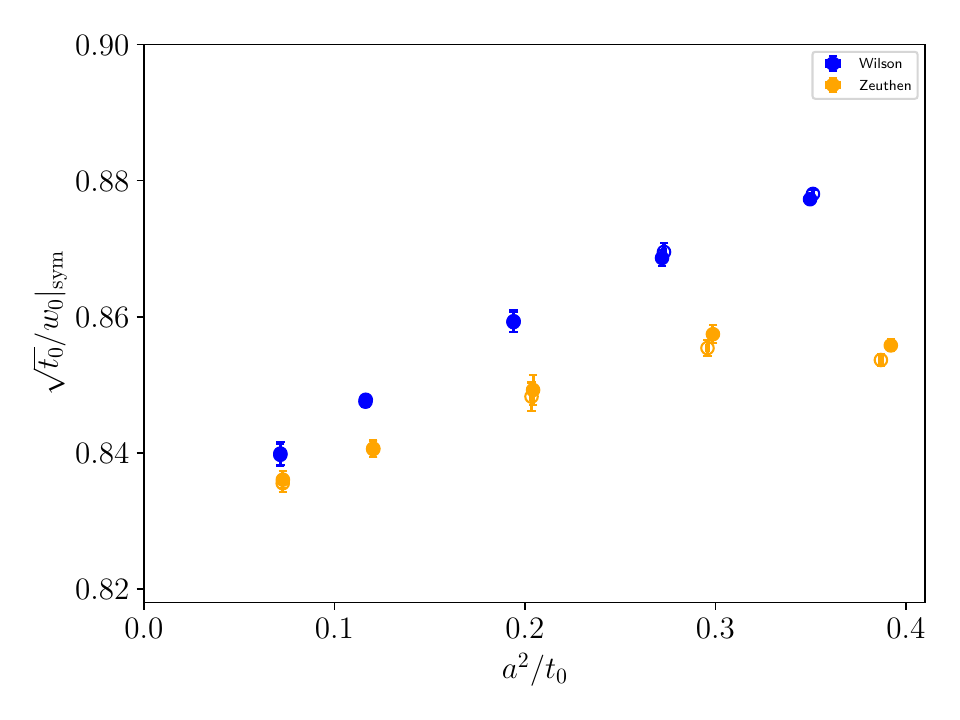}
  \includegraphics[width=.49\textwidth]{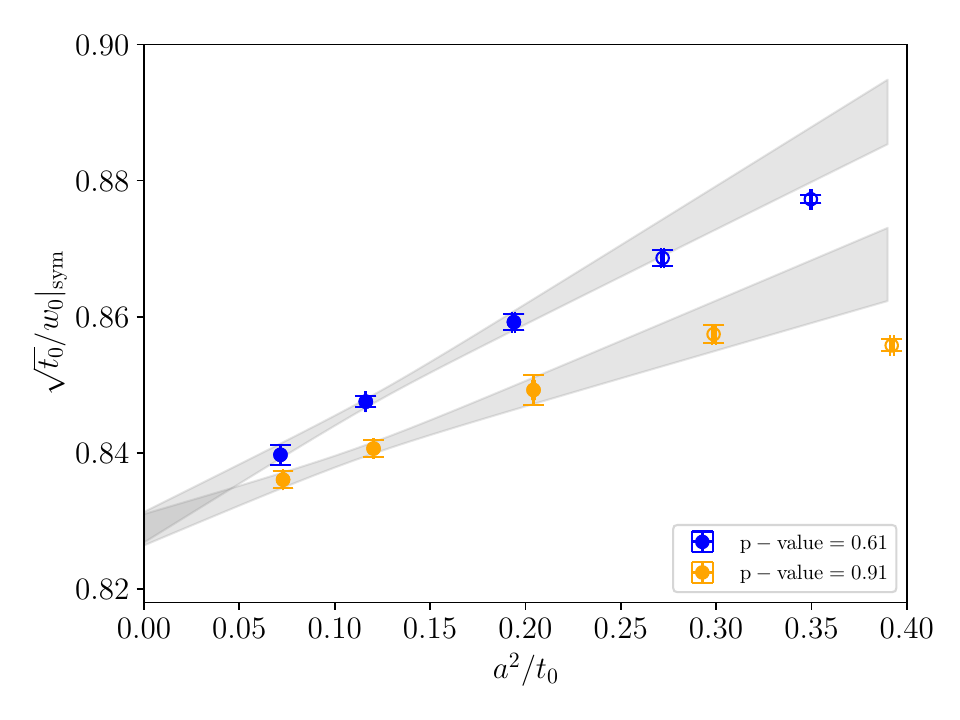}
  \caption{\label{fig:w0t0_shift}
  \textit{Left:} Comparison of $\sqrt{t_0}/w_0$ for the symmetric ensembles before (empty points) and after (filled points) the shift to $\phi_2^{\rm sym}=0.74$, employing \cref{eq:md}, both for Wilson and Zeuthen flow.
  Blue points are for Wilson-flow data, while orange points are for Zeuthen-flow data.
  The horizontal coordinate $a^2/t_0$ uses the Wilson or Zeuthen determination of $t_0$ corresponding to each point.
  \textit{Right:} Continuum limit extrapolation of $\sqrt{t_0}/w_0$ for Wilson- and Zeuthen-flow data.
  Empty symbols are not included in the fit, and a common continuum limit is not imposed.
  }
\end{figure}

We define the SU(3)-symmetric line of constant physics by $m_{\rm ud}=m_{\rm s}$ and $\phi_2^{\rm sym}=0.74$.\footnote{The final results at the physical point are independent of this particular choice.}
The SU(3)-symmetric ensembles on the ${\rm tr}(M_q)={\rm cnst.}$ trajectory, with $m_\pi=m_K\approx420$ MeV (cf. \ret{tab:ens}), lie close to this line but have small mistunings in $\phi_2$.
To shift them to the target value and impose the line of constant physics as we approach the continuum limit, we use a first-order Taylor expansion in the quark masses,
\begin{equation}
  \label{eq:md}
\mathcal{O}(m_q')=\mathcal{O}(m_q)+\sum_q\delta m_q\left.\frac{d\mathcal{O}}{dm_q}\right|_{m_q},
\end{equation}
where $d\mathcal{O}/dm_q$ are the measured derivatives of observable $\mathcal{O}$ with respect to the simulated quark masses $m_q$ (with contributions both from the sea and valence sectors, cf. \cite{Bruno:2016plf,Ce:2022kxy,Bussone:2025wlf}), $\delta m_q\equiv m_q'-m_q$, and we tune $\delta m_q$ to the needed value in order to impose the LCP condition $\phi_2^{\rm sym}=0.74$.
This procedure is applied to the lattice observables $\mathcal{O}=\{t_0/a^2, w_0/a, am_{\pi}, am_K, af_{\pi}\}$, and it determines $\sqrt{t_0}/w_0|_{\rm sym}$ for each value of $\beta$.
The required value of $\delta m_q$ is slightly different for the Wilson- and Zeuthen-flow data, as $t_0/a^2$ from both types of flow is used in the definition of $\phi_2$.
In the left panel of \refig{fig:w0t0_shift} we show the values of $\sqrt{t_0}/w_0$ for the symmetric ensembles before and after the shift to $\phi_2^{\rm sym}=0.74$, while in Tables \ref{tab:obs_sym_wilson}--\ref{tab:obs_sym_zeuthen} we show the corresponding numerical values.
Note that for $\beta=3.46,3.70$ the ensembles used in the Wilson- and Zeuthen-flow cases are not the same -- due to data availability -- but all are symmetric point ensembles (cf. \ret{tab:ens}).
In the right panel of \refig{fig:w0t0_shift}, we show separate continuum extrapolations of the Wilson- and Zeuthen-flow data after applying the shift, using only the finer lattice spacings and without imposing a common continuum limit. The extrapolated values are in good agreement.
\begin{table}[htbp]
\centering
\begin{minipage}[t]{0.48\textwidth}
\centering
\begin{tabular}{c c c}
\toprule
id & $\beta$ & $\sqrt{t_0}/w_0|_{\rm sym}$ [Wilson flow] \\
\midrule
H101 & 3.40 & 0.8773(6) \\
\midrule
B450 & 3.46 & 0.8686(12) \\
\midrule
N202 & 3.55 &  0.8592(15) \\
\midrule
J307 & 3.70 & 0.8475(7) \\
\midrule
J500 & 3.85 & 0.8397(17) \\
\bottomrule
\end{tabular}
\caption{\label{tab:obs_sym_wilson}%
Results of $\sqrt{t_0}/w_0|_{\rm sym}$ for the Wilson flow, determined by shifting the symmetric point ensembles to $\phi_2^{\rm sym}=0.74$ with the use of \cref{eq:md} and the Wilson-flow data for $t_0/a^2$.%
}
\end{minipage}
\hfill
\begin{minipage}[t]{0.48\textwidth}
\centering
\begin{tabular}{c c c}
\toprule
id & $\beta$ & $\sqrt{t_0}/w_0|_{\rm sym}$ [Zeuthen flow] \\
\midrule
H101 & 3.40 & 0.8558(9) \\
\midrule
H400 & 3.46 & 0.8574(13) \\
\midrule
N202 & 3.55 & 0.8492(22) \\
\midrule
N300 & 3.70 & 0.8406(12) \\
\midrule
J500 & 3.85 & 0.8361(13) \\
\bottomrule
\end{tabular}
\caption{\label{tab:obs_sym_zeuthen}%
Results of $\sqrt{t_0}/w_0|_{\rm sym}$ for the Zeuthen flow, determined by shifting the symmetric point ensembles to $\phi_2^{\rm sym}=0.74$ with the use of \cref{eq:md} and the Zeuthen-flow data for $t_0/a^2$.%
}
\end{minipage}
\end{table}
\subsection{\textit{Higher-order cutoff effects}}
As will be seen in the next subsection, at fine lattice spacing, our data for $\sqrt{t_0}/w_0|_{\rm sym}$ can be properly described by a linear behaviour in $a^2$.
Thus, in order to asses for higher-order cutoff effects without overfitting, instead of trying to fit an $a^3$ or $a^4$ term, we proceed as follows.
The direct ratio $\sqrt{t_0}/w_0|_{\rm sym}$ can be expressed as
\begin{equation}
  \label{eq:fun1}
\left.\frac{\sqrt{t_0}}{w_0}\right|_{\rm sym}=A+B\frac{a^2}{t_0^{\rm sym}}.
\end{equation}
The inverse ratio $w_0/\sqrt{t_0}|_{\rm sym}$ can be expressed in the same way, albeit with different coefficients
\begin{equation}
  \label{eq:fun2}
\left.\frac{w_0}{\sqrt{t_0}}\right|_{\rm sym}=A'+B'\frac{a^2}{t_0^{\rm sym}}.
\end{equation}
After fitting and expressing both extrapolations as $\sqrt{t_0}/w_0$, their continuum estimates agree within one standard deviation and have comparable fit qualities.
Yet their difference is a $\mbox{O}(a^4)$ cutoff effect.
The latter is responsible for a small systematic effect in the continuum value of $\sqrt{t_0}/w_0|_{\rm sym}$, depending on whether the direct or inverse ratio is extrapolated.
Thus, including both types of fit in a model average, as will be shown below, allows to assess the size of $\mbox{O}(a^4)$ cutoff effects.
This treatment of higher-order cutoff effects takes into account only even powers in $a$, but neglects odd ones.
For pure Yang--Mills theory, only even powers are expected for
bulk flow observables \cite{Ramos:2015baa}.
With dynamical Wilson fermions -- like the CLS ensembles considered
in this work -- odd powers may arise through sea-quark effects.
These may be perturbatively suppressed, but their size for the
present observable is not known.
We account for such effect by performing cuts in the lattice spacing and including the resulting variation in the continuum limit in our systematic uncertainty via a model average.
\subsection{\textit{Model average}}
We will assess and incorporate the dependence of our fit results on variations of the fit ansatz and cuts in the data with a model averaging procedure.
To do so, we may use an Information Criterion (IC) argument, which assigns to each $i$-model a weight according to
\begin{equation}
	\label{eq:W}
	W_i\propto\exp\left(-\frac{1}{2}{\rm IC}_i\right),
\end{equation}
where weights are normalized to unity $\sum_iW_i=1$, and ${\rm IC}_i$ is the information criterion of a given $i$-model.
Two popular IC choices are the Takeuchi Information Criterion (TIC), given by \cite{Frison:2023lwb}
\begin{equation}
	\label{eq:TIC}
	{\rm TIC}=\chi^2-2\left<\chi^2\right>,
\end{equation}
where $\left<\chi^2\right>$ is defined in \cite{Bruno:2022mfy}, and the (modified) Akaike Information Criterion ($\rm AIC^{\rm sub}$, e.g. \cite{ethan_new})
\begin{equation}
	\label{eq:AIC}
	{\rm AIC^{\rm sub}}=\chi^2+2n_{\rm param}+n_{\rm cut},
\end{equation}
with $n_{\rm param}$ the number of fit parameters and $n_{\rm cut}$ the number of data points removed from the fit.
The model average result for observable $\mathcal{O}$, referring to each $i$-model's result by $\mathcal{O}_i$, is then given by
\begin{equation}
	\label{eq:Oav}
	\left<\mathcal{O}\right>=\sum_i\mathcal{O}_iW_i,
\end{equation}
and a systematic uncertainty can be assigned as
\begin{equation}
	\label{eq:syst}
	\sigma_{\rm syst}^2=\left<\mathcal{O}^2\right>-\left<\mathcal{O}\right>^2.
\end{equation}
An alternative way of quoting a model averaged result is the following:
we (i) fit to a constant the different $\{\mathcal{O}_i\}$ to define the model average central value and statistical uncertainty; then (ii) define the total -- systematic and statistical -- uncertainty as the maximum spread between the various $\{\mathcal{O}_i\}$, and (iii) compute the systematic uncertainty from (ii) and (i) by considering that systematic and statistical uncertainties add up in quadrature.
In all this procedure, we only take into account models with good p-values -- as computed with the definition in \cite{Bruno:2022mfy} --, conventionally defining the latter as ``good'' if p-value$\,\geq0.1$.%
\footnote{Our result in \cref{eq:continuum} remains unchanged, albeit with a slight reduction in the systematic uncertainty, if additionally one imposes the cut p-value$\,\leq0.9$ in order to avoid overfitting.}
We refer to this method as the ``MDE'' or ``Maximum Deviation Envelope''.
By construction, this method covers all ``good'' models within approximately one standard deviation.
In the case where the spread of fit results is large with respect to the statistical uncertainties, we consider the MDE method to provide a better, more conservative, estimate of the size of the systematic uncertainty, as illustrated in \refig{fig:continuum_av}.
\subsection{\textit{Continuum limit extrapolation and model average}}
Having defined an LCP by the $\phi_2^{\rm sym}=0.74$ condition, we perform the continuum limit extrapolation of $\sqrt{t_0}/w_0|_{\rm sym}$.
To incorporate the effect of non-zero anomalous dimensions that modify the leading scaling behavior in the Symanzik expansion, we explore including $\mbox{O}(a^2\alpha_s^{\Gamma})$ cutoff effects.
For this, we choose $\Gamma\in\{0.0,+0.760\}$, corresponding to the minimum possible value of $\Gamma$ for our action and the expected dominant contribution, respectively \cite{Husung:2021mfl}.

\begin{figure}[t!]
  \centering
  \includegraphics[width=.49\textwidth]{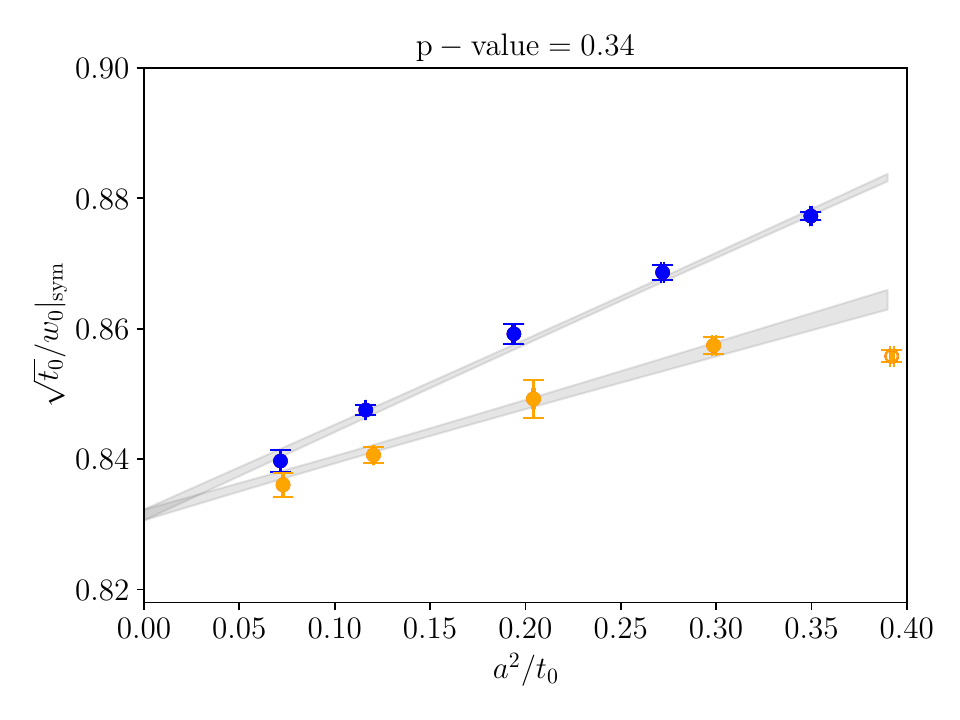}
  \includegraphics[width=.49\textwidth]{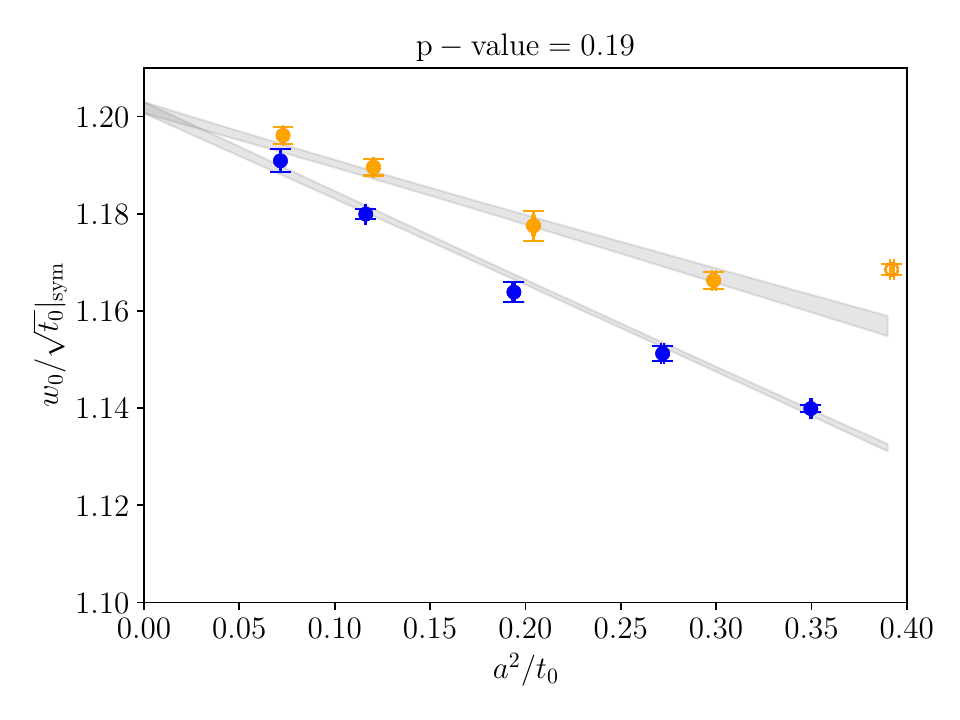}
  \includegraphics[width=.49\textwidth]{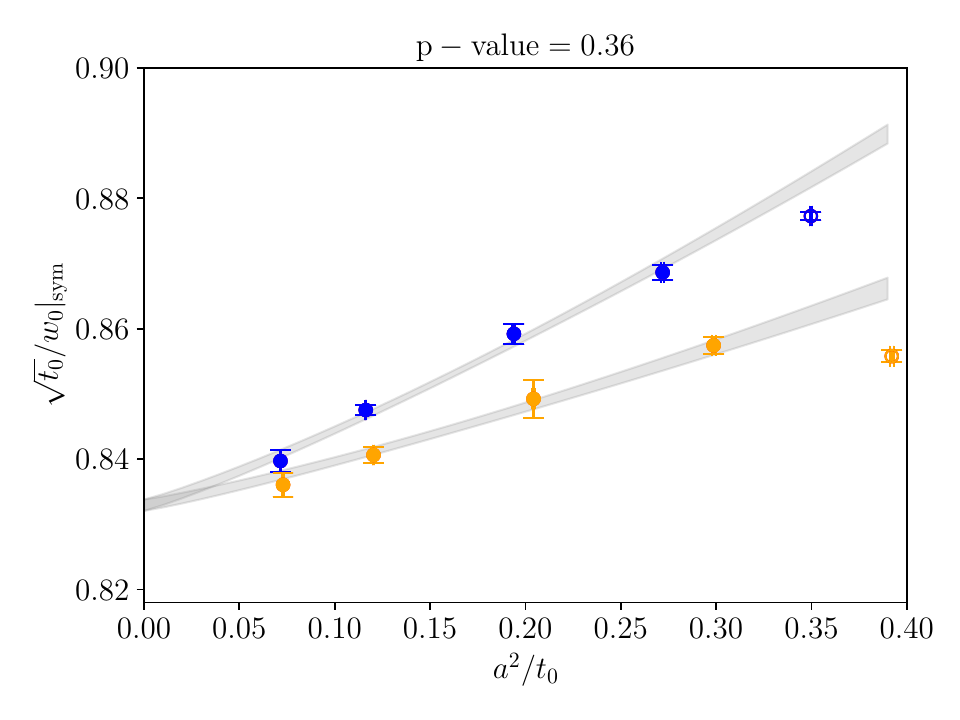}
  \includegraphics[width=.49\textwidth]{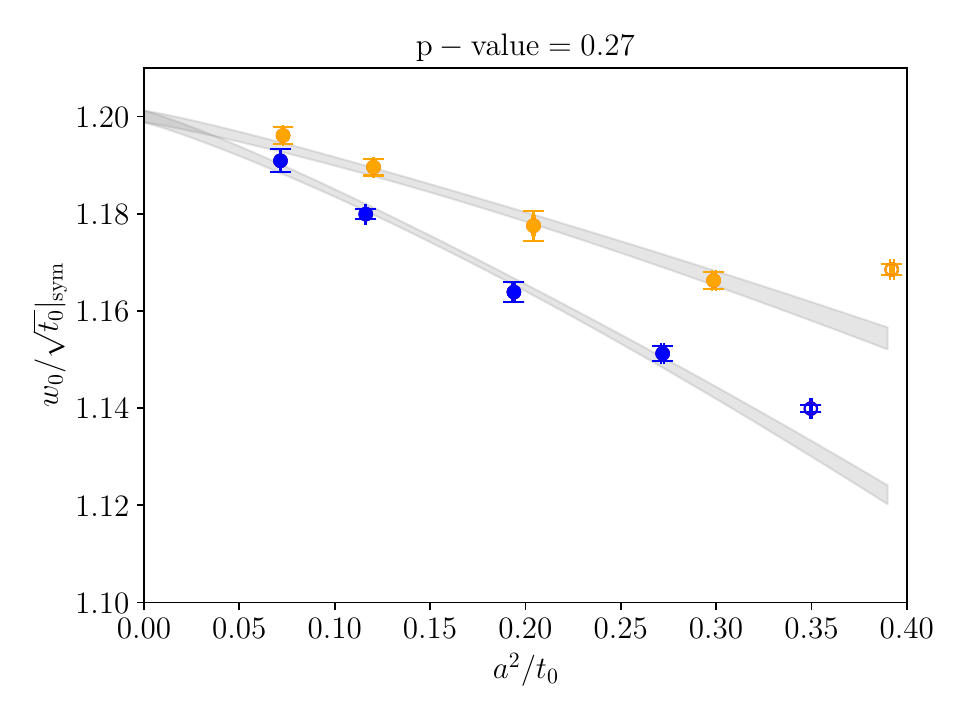}
  \caption{\label{fig:continuum}
  \textit{Top row}: fit of $\sqrt{t_0}/w_0|_{\rm sym}$ (left) and $w_0/\sqrt{t_0}|_{\rm sym}$ (right) to \cref{eq:continuum_fit_function}, setting $\Gamma=0.0$.
  \textit{Bottom row}: the same as top row, setting $\Gamma=0.760$.
  Empty points are not included in the fit.
  Blue points are Wilson-flow data, while orange points are Zeuthen-flow data.
  }
\end{figure}

\begin{figure}[t!]
  \centering
  \includegraphics[width=.8\textwidth]{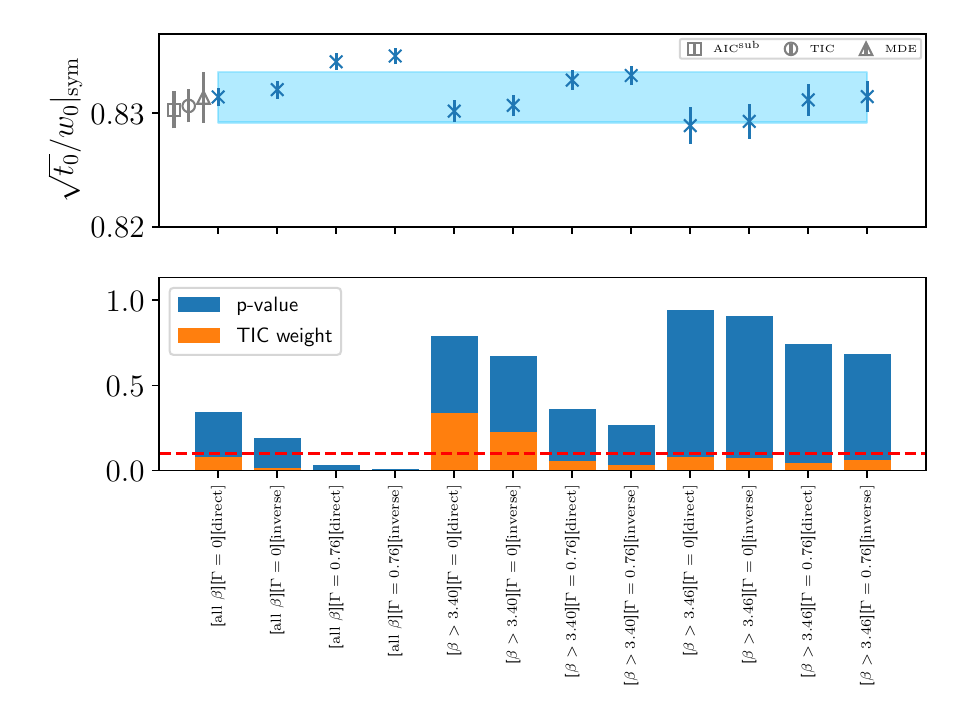}
  \caption{\label{fig:continuum_av}
  Model exploration and model average (blue horizontal band) of the continuum extrapolation of $\sqrt{t_0}/w_0|_{\rm sym}$.
  In the top plot we provide with the model average results obtained from the $\rm AIC^{\rm sub}$ (gray square point), TIC (gray circle) and MDE (gray triangle), choosing the MDE as our final estimate, which gives the horizontal blue band.
  In the bottom plot we show the p-values of each individual model, the red horizontal dashed line marking the (conventional) p-value$\,\geq0.1$ threshold.
  }
\end{figure}

In addition, as introduced in the previous subsection, we explore fiting the inverse of the ratio -- i.e. $w_0/\sqrt{t_0}$ -- to the same fit function.
This way the fit function for both $\sqrt{t_0}/w_0$ and $w_0/\sqrt{t_0}$ reads
\begin{equation}
  \label{eq:continuum_fit_function}
F=A+B\frac{a^2}{t_0^{\rm sym}}\alpha_s(1/a)^{\Gamma},
\end{equation}
where we use $\alpha_s(1/a)\propto-1/\log(a\Lambda_{\rm QCD})$.
The variation between the continuum limit of $\sqrt{t_0}/w_0$ and that of $w_0/\sqrt{t_0}$ is a measure of $\mbox{O}(a^4)$ cutoff effects.
In addition to these variations, we also perform cuts in the data: removing the coarsest lattice spacing $\beta=3.40$ or the two coarsest $\beta\leq3.46$, while we always remove $\beta=3.40$ for the Zeuthen data as it clearly deviates for a linear behaviour in $a^2$ (cf. \refig{fig:continuum}).\footnote{
Note that although the Zeuthen flow removes the classical $\mbox{O}(a^2)$ cutoff effects, these could have opposite sign to dynamically generated $\mbox{O}(a^2)$ cutoff effects, such that overall, Zeuthen flow may exhibit stronger $\mbox{O}(a^2)$ cutoff effects.
This is not the case in our data set (cf. \refig{fig:continuum}), though stronger higher-order effects are seen for the coarsest lattice spacing for the Zeuthen-flow data.
}
A common continuum limit between the Zeuthen- and Wilson-flow data is imposed.
In \refig{fig:continuum} we illustrate the continuum limit extrapolation for some of these models.
Whereas the deviation from the continuum limit is reduced for Zeuthen-flow data, indications of the onset of higher-order lattice artifacts can be seen at the coarsest value of the lattice spacing considered in this work.
The combined extrapolation of the data does not pose any difficulty. 
Whereas fits to either set of data gives very similar results, we find the combined fits to be less sensitive to fluctuations in the data.
In \refig{fig:continuum_av} we give an overview of the results of the twelve fits that are performed for this continuum extrapolation.
The model labels in the horizontal axis of  are organized as follows:
\begin{itemize}
	\item \textbf{1st tag:} $[\rm all\;\beta]$, $[\beta>3.40]$ or $[\beta>3.46]$ indicates which values of $\beta$ are included in the fit, assuming $\beta=3.40$ for Zeuthen-flow data is always removed from fits.
	\item \textbf{2nd tag:} $[\Gamma=0]$ or $[\Gamma=0.76]$ indicates the value of $\Gamma$ used in \cref{eq:continuum_fit_function}.
	\item \textbf{3rd tag:} [direct] or [inverse] indicates whether the ratio $\sqrt{t_0}/w_0|_{\rm sym}$ or its inverse, respectively, is fitted to \cref{eq:continuum_fit_function}.
\end{itemize}

The upper panel shows the individual results together with the model averages from the three criteria introduced above. 
The two gray symbols to the left denote the results of the $\rm AIC^{\rm sub}$ and the TIC, respectively.
The lower panel provides the TIC weights and the p-values of the fits, respectively.

As can be seen in the upper panel, due to the high statistical accuracy and the relatively strong cutoff effects, we find a significant spread of the results that is not covered by either of the two information criteria.
We therefore opt to base the final estimate from the scan of fit models on the result from the more conservative MDE method which is denoted by the blue horizontal band.
The corresponding result for $\sqrt{t_0}/w_0|_{\rm sym}$ in the continuum reads
\begin{equation}
  \label{eq:continuum}
\left.\frac{\sqrt{t_0}}{w_0}\right|_{{\rm sym, \; cont}}=0.8314(10)_{\rm stat}(20)_{\rm cont}[22]_{\rm total},
\end{equation}
where the first error refers to the statistical uncertainty coming from the gauge fluctuations, and the second refers to the systematic uncertainty in the continuum limit extrapolation as estimated by the MDE method above.

\section{Chiral interpolation}
\label{sec:chiral_extrap}

In this section we perform the chiral interpolation -- and residual continuum extrapolation -- of $(\sqrt{t_0}/w_0)/(\sqrt{t_0}/w_0|_{\rm sym})$.
In this ratio of ratios most of the cutoff effects are expected to cancel, except for residual mass-dependent effects.
For this analysis, in which our full set of ensembles in \ret{tab:ens} is used, we rely on Wilson-flow data only due to data availability.

Above, we were free to choose the exact definition of the SU(3)-symmetric point such that the mistuning of the gauge ensembles with respect to our definition was small, especially at fine lattice spacing.
Since the CLS ensembles have been generated based on a tuning that relied on $\phi_2$, it was natural to use $\phi_2$ as quark mass proxy for the first step, in whose definition appears $t_0$.
However, the physical point targeted in this work is defined by the Edinburgh consensus in \cref{eq:edinburgh}.
Therefore, we will now switch to the $\rho$-variables defined in \cref{eq:rhos} as kinematical variables.
For the chiral interpolation of the data, we start with the NNLO SU(3) $\chi$PT prediction for the flow scales $t_0$ and $w_0$ \cite{Bar:2013ora}, which we write in the form
\begin{align}
	\label{eq:ChPT_t0}
	t_0
	&= t_{0,\rm ch}\Biggl[
	1
	+\frac{2k_1}{8t_0(4\pi f)^2}
	\left(\phi_K+\frac{1}{2}\phi_\pi\right)
	+\frac{3k_2-k_1}{(8t_0)^2(4\pi f)^4}
	\phi_\pi^2
	\log\left(\frac{\phi_\pi}{8t_0\Lambda_\chi^2}\right)
	\nonumber\\
	&\quad
	+\frac{4k_2}{(8t_0)^2(4\pi f)^4}
	\phi_K^2
	\log\left(\frac{\phi_K}{8t_0\Lambda_\chi^2}\right)
	+\frac{k_2-k_1}{(8t_0)^2(4\pi f)^4}
	\phi_\eta^2
	\log\left(\frac{\phi_\eta}{8t_0\Lambda_\chi^2}\right)
	\nonumber\\
	&\quad
	+\frac{4k_4}{(8t_0)^2(4\pi f)^4}
	\left(\phi_K+\frac{1}{2}\phi_\pi\right)^2
	+\frac{k_5}{(8t_0)^2(4\pi f)^4}
	\left(\phi_K-\phi_\pi\right)^2
	\Biggr].
\end{align}
with $\phi_P=8t_0m_P^2$, and the Gell-Mann-Okubo relation $m_{\eta}^2=\frac{4}{3}m_K^2-\frac{1}{3}m_{\pi}^2$.
The expression for $w_0$ is identical, albeit with different $k$-coefficients.
Writing these expressions in terms of dimensionless quantities like $\phi_P$ allows for a straightforward connection with the lattice implementation.
In practice, we will use the $\rho$-variables instead of the $\phi$-variables for the chiral interpolation, which can be done by setting $\Lambda_{\chi}=4\pi f$ and noting
\begin{equation}
\frac{\phi_P}{8t_0f^2}=\frac{m_P^2}{f^2}\underset{f\rightarrow f_{\pi}}{\longrightarrow}\rho_P,
\end{equation}
Replacing $f\rightarrow f_{\pi}$ is a higher-order effect which can be reabsorbed in the values of the $k$-coefficients.
Combining the above replacements, the $\chi$PT-inspired fit function for $\sqrt{t_0}/w_0$ reads
\begin{align}
	\label{eq:ChPT}
	\frac{\sqrt{t_0}}{w_0}
	&= A\Biggl[
	1
	+\frac{2C}{(4\pi)^2}\rho_4
	+\frac{3D-C}{(4\pi)^4}\rho_2^2
	\log\left(\frac{\rho_2}{(4\pi)^2}\right)
	+\frac{4D}{(4\pi)^4}
	\left(\rho_4-\frac{1}{2}\rho_2\right)^2
	\log\left(
	\frac{\rho_4-\frac{1}{2}\rho_2}{(4\pi)^2}
	\right)
	\nonumber\\
	&\quad
	+\frac{D-C}{(4\pi)^4}
	\left(\frac{4}{3}\rho_4-\rho_2\right)^2
	\log\left(
	\frac{\frac{4}{3}\rho_4-\rho_2}{(4\pi)^2}
	\right)
	+\frac{4E}{(4\pi)^4}\rho_4^2
	+\frac{F}{(4\pi)^4}
	\left(\rho_4-\frac{3}{2}\rho_2\right)^2
	\Biggr].
\end{align}
Here, we set the fit parameter $E$ to zero since we cannot resolve this term with the current data where the variation of $\rho_4$ independent of a variation of $\rho_2$, as realized on the $m_{\rm s}\approx m_{\rm s}^{\rm ph}$ trajectory, is small, cf.~\refig{fig:rho_phi}.
Because we can interpolate to physical quark masses, we do not have to rely on the guidance of chiral perturbation theory.
Since our data extents up to pion masses of about 420\,MeV, it is not clear that the effective theory up to NNLO can describe the data sufficiently well.
To rule out that our results at the physical point are affected by a biasing effect, we also explore a polynomical expression in the pion and kaon masses,\footnote{If instead we Taylor expand \cref{eq:ChPT} around the symmetric point, i.e. $\sqrt{t_0}/w_0=A+B\rho_4+C\rho_4^2+D\left(\rho_4-\frac{3}{2}\rho_2\right)^2$, we find virtually no difference with respect to \cref{eq:Tay} in the fitting result.}
\begin{equation}
  \label{eq:Tay}
\frac{\sqrt{t_0}}{w_0}=A+B\rho_2+C\rho_2^2+D\rho_4.
\end{equation}
Dividing both expressions in \cref{eq:ChPT,eq:Tay} by the corresponding expression evaluated at the LCP given by $\phi_2^{\rm sym}=0.74$ (and equivalently $\phi_4^{\rm sym}=\frac{3}{2}\phi_2^{\rm sym}=1.11$) defines our set of fit functions for the $(\sqrt{t_0}/w_0)/(\sqrt{t_0}/w_0|_{\rm sym})$ ratio.
For this ratio of ratios, we parameterize the leading residual
mass-dependent cutoff effects as
\begin{equation}
		\label{eq:cutoff_chiral}
		\begin{aligned}
			\left(
			\frac{\sqrt{t_0}/w_0}
			{(\sqrt{t_0}/w_0)_{\rm sym}}
			\right)^{\rm latt}
			={}&
			\left(
			\frac{\sqrt{t_0}/w_0}
			{(\sqrt{t_0}/w_0)_{\rm sym}}
			\right)^{\rm cont}
			\\
			&+\frac{a^2}{t_0^{\rm sym}}
			\left[
			c_\Delta\left(\rho_4-\frac{3}{2}\rho_2\right)^2
			+c_{\rm tr}\left(
			\frac{t_0^{\rm sym}}{t_0}\phi_4-1.11
			\right)
			\right].
		\end{aligned}
	\end{equation}
Here $c_{\Delta}$ parameterizes cutoff effects associated with flavor
	splitting, while $c_{\rm tr}$ allows for a dependence on the singlet-mass
	proxy $m_K^2+\frac{1}{2}m_\pi^2$ away from the symmetric point.
Note that since the LCP was defined with the $\phi$- instead of the $\rho$-variables used here, the values of $\rho_2^{\rm sym}$ do not coincide for different choices of $\beta$,\footnote{Rather it is $\phi_2^{\rm sym}=0.74$ what is kept constant for different values of $\beta$.} the difference being a cutoff effect.
\begin{figure}[t!]
	\centering
	\includegraphics[width=.8\textwidth]{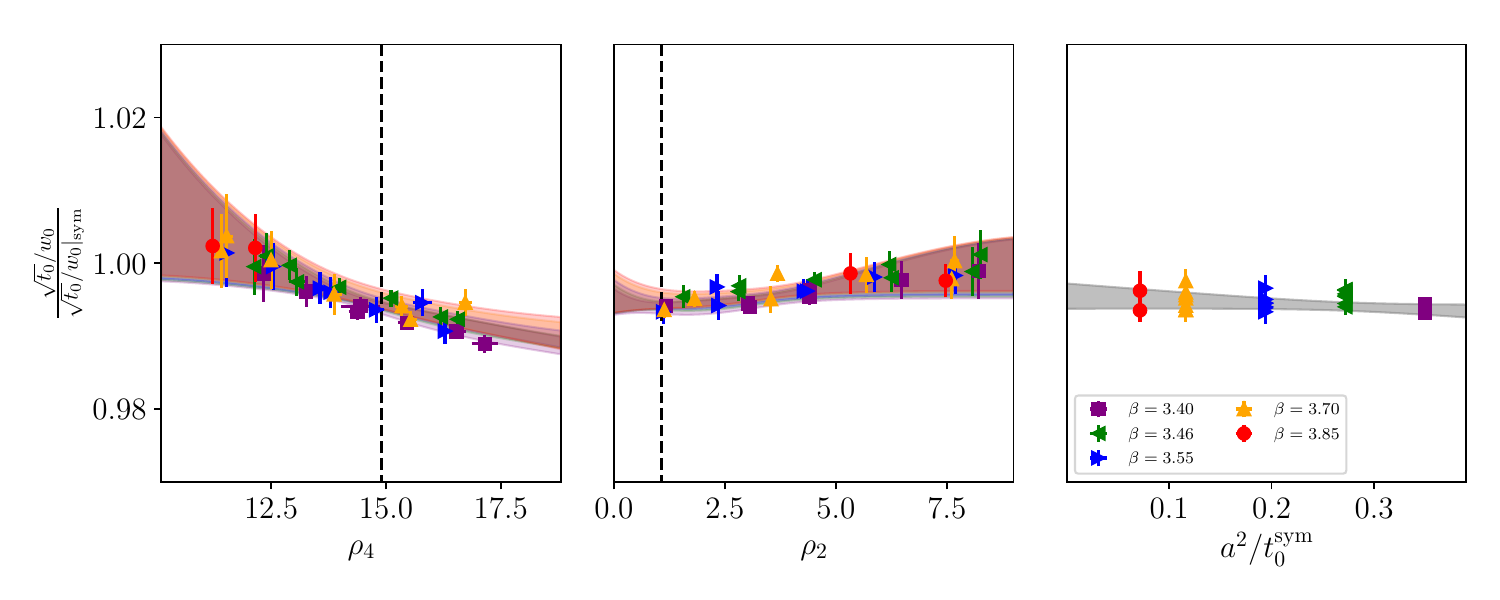}
	\includegraphics[width=.8\textwidth]{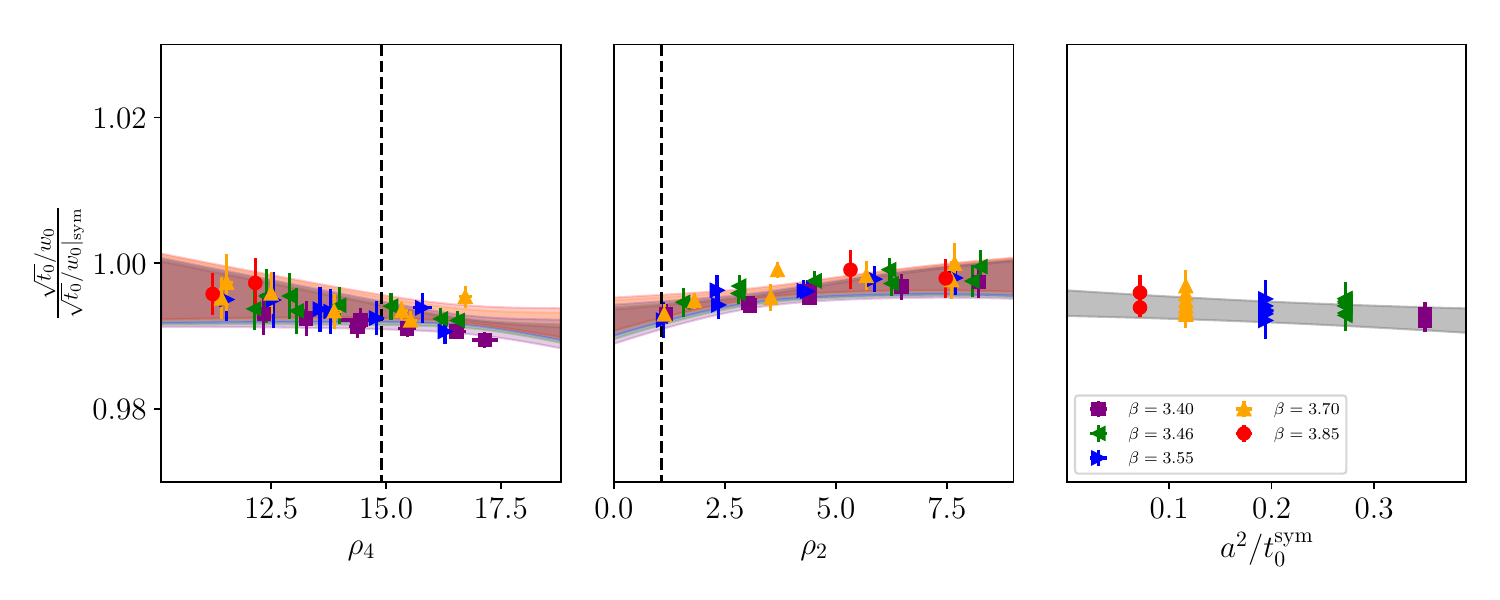}
	\caption{\label{fig:chiral}
		\textit{Top panel:} Chiral interpolation and residual continuum extrapolation of $(\sqrt{t_0}/w_0)/(\sqrt{t_0}/w_0|_{\rm sym})$ employing a NNLO SU(3) $\chi$PT-inspired ansatz as in \cref{eq:ChPT}.
		\textit{Bottom panel:} The same, employing a Taylor expansion as in \cref{eq:Tay}.
		In the left plots, points were projected to $\rho_2=\rho_2^{\rm ph}$, in the middle plots to $\rho_4=\rho_4^{\rm ph}$, and in the right plots to both $\rho_2=\rho_2^{\rm ph}$ and $\rho_4=\rho_4^{\rm ph}$.
		The vertical dashed lines mark the physical point.
	    As in all plots at least one $\rho_i$ is projected to its physical value, the symmetric point is not shown here, and thus no point lies at unity.
	}
\end{figure}
In addition to this variation of the fit ansatz, we again explore performing cuts in the data: We perform fits where we exclude the coarsest lattice spacing at $\beta=3.40$ and the two coarsest lattice spacings, $\beta\leq3.46$. 
We also include a cut in the pion mass, $m_{\pi}\geq340$ MeV.
We show some representative models in \refig{fig:chiral}.

Several observations can be made. 
First, the overall effect of departing from the SU(3)-symmetric point is small, at most 1\% on the ensembles considered in this work.
Second, as can be seen from the comparison of the upper and lower panel, the variation of the fit forms allows quite some freedom in describing the data. 
However, the mass dependence is well constrained in the vicinity of the physical point, denoted by the dashed vertical lines.
Cutoff effects are mild, but still noticeable.

\begin{figure}[t!]
  \centering
  \includegraphics[width=.8\textwidth]{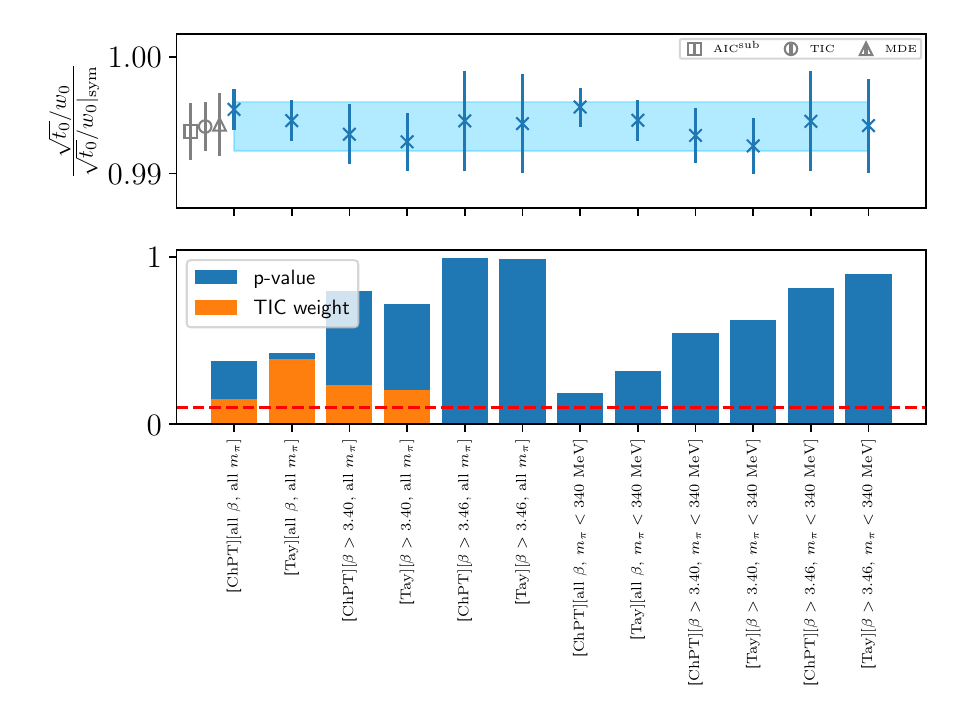}
  \caption{\label{fig:chiral_av}
  Model exploration and model average (blue horizontal band) of the chiral interpolation and residual continuum extrapolation of $(\sqrt{t_0}/w_0)/(\sqrt{t_0}/w_0|_{\rm sym})$.
  In the top plot we provide with the model average results obtained from the $\rm AIC^{\rm sub}$ (gray square point), TIC (gray circle) and MDE (gray triangle), choosing the TIC as our final estimate, which gives the horizontal blue band.
  In the bottom plot we show the weights of each of them as given by \cref{eq:W}, and corresponding p-values.
  The red horizontal dashed line marks the p-value$\,\geq0.1$ conventional threshold.
  }
\end{figure}

The upwards bending in the top pannel with respect to the bottom one is due to chiral logarithms, absent in the Taylor fits.
However, as noted above, the inclusion of physical pion ensembles that allow for a chiral interpolation guarantee that around the physical point both types of fit ansätze give very similar results.
\refig{fig:chiral_av} gives an overview of the fits that enter the model average to determine the ratio of ratios at the physical point, where the layout is the same as in \refig{fig:continuum_av}.
The model labels in the horizontal axis are organized as follows:
\begin{itemize}
\item \textbf{1st tag:} [ChPT] or [Tay] indicates whether \cref{eq:ChPT} or \cref{eq:Tay} is used for the chiral interpolation.
\item \textbf{2nd tag:} indicates the set of $\beta$ and/or $m_{\pi}$ values included in the fit.
\end{itemize}
With respect to the statistical uncertainties, the spread of the results is mild. 
We note that the effect of changing the functional form has a similar size as the effect from excluding the coarsest lattice spacing.
A cut in the pion mass has no noticeable effect on the result at the physical point.
Both information criteria give similar results and the systematic uncertainties cover well the dispersion of the individual models.
We base our final result on the TIC (cf. \cref{eq:TIC}), providing
\begin{equation}
  \label{eq:chiral}
\left.\frac{\sqrt{t_0}/w_0}{\sqrt{t_0}/w_0|_{\rm sym}}\right|_{\rm ph}=0.9940(19)_{\rm stat}(9)_{\rm \chi-cont}[21]_{\rm total}\,,
\end{equation}
and note that all individual results entering this average differ at most by one standard deviation from this result.
The effect of deviating from the SU(3)-symmetric point to physical quark masses is thus below one percent and we determine it with a relative precision of 0.20\%.

Combining this result with the result in \cref{eq:continuum}, we obtain
\begin{equation}
  \label{eq:res}
\left.\frac{\sqrt{t_0}}{w_0}\right|_{\rm ph}=0.8264(15)_{\rm stat}(20)_{\rm cont}(8)_\text{$\chi$-cont}[26]_{\rm total},
\end{equation}
where ``cont'' refers to the systematic uncertainty of the continuum extrapolation of $\sqrt{t_0}/w_0|_{\rm sym}$ in \cref{eq:continuum}, and ``$\chi$-cont'' to that of the chiral interpolation and residual continuum extrapolation of $(\sqrt{t_0}/w_0)/(\sqrt{t_0}/w_0|_{\rm sym})$ in \cref{eq:chiral}.
For the definition of the physical point we used the scheme defined by the Edinburgh consensus \cite{FlavourLatticeAveragingGroupFLAG:2024oxs} in \cref{eq:edinburgh}.
The automatic differentiation tools as implemented in the \verb|ADerrors.jl| package \cite{Ramos:2018vgu} allow us to compute the derivatives of our result in \cref{eq:res} with respect to the inputs used in said scheme, which are quoted in \ret{tab:scheme_der}.
\begin{table}[tbp]
\centering
\begin{tabular}{ccc}
\toprule
$\displaystyle\left.\frac{\partial(\sqrt{t_0}/w_0)}{\partial m_{\pi}}\frac{m_{\pi}}{\sqrt{t_0}/w_0}\right|_{\rm ph}$ &
$\displaystyle\left.\frac{\partial(\sqrt{t_0}/w_0)}{\partial m_{K}}\frac{m_{K}}{\sqrt{t_0}/w_0}\right|_{\rm ph}$ &
$\displaystyle\left.\frac{\partial(\sqrt{t_0}/w_0)}{\partial f_{\pi}}\frac{f_{\pi}}{\sqrt{t_0}/w_0}\right|_{\rm ph}$ \\
\midrule
$5.099 \times 10^{-4}$ & $-2.687 \times 10^{-2}$ & $2.636 \times 10^{-2}$ \\
\bottomrule
\end{tabular}
\caption{\label{tab:scheme_der} Derivatives of $\sqrt{t_0}/w_0$ at the physical point in \cref{eq:res} with respect to the inputs used in the Edinburgh consensus scheme (cf. \cref{eq:edinburgh}).}
\end{table}
The ratio of flow scales at the physical point allows to relate existing determinations of either flow scale to each other.
Two results for the flow scale $\sqrt{t_0^{\rm ph}}$, based on a similar set of CLS ensembles as used in this work, were obtained in \cite{Bussone:2025wlf}.
One is based on the Edinburgh consensus scheme (cf. \cref{eq:edinburgh}) and the other on a mixed scheme, combining the inputs in \cref{eq:edinburgh} with the FLAG average for $f_K^{\rm isoQCD}=157.4(6)$ \cite{FlavourLatticeAveragingGroupFLAG:2024oxs}.
Using said determinations, together with our result for the ratio $\sqrt{t_0}/w_0|_{\rm ph}$, allows us to extract
\begin{align}
  \label{eq:w0_ph_mix}
w_0^{\rm ph}&=0.1743(7)_{\rm stat}(4)_{\rm cont}(2)_\text{$\chi$-cont}(3)_{t_0^{\rm ph}}[9]_{\rm total}\;{\rm fm}, \quad\quad\quad {\rm mixed\; scheme}, \\
\label{eq:w0_ph}
w_0^{\rm ph}&=0.1742(11)_{\rm stat}(4)_{\rm cont}(2)_\text{$\chi$-cont}(11)_{t_0^{\rm ph}}[17]_{\rm total}\;{\rm fm}, \quad {\rm Edinburgh\; scheme},
\end{align}
where the ``$t_0^{\rm ph}$'' uncertainty comes from the systematic uncertainty in the value of $\sqrt{t_0^{\rm ph}}$ from \cite{Bussone:2025wlf}, while the statistical uncertainty of this latter quantity is correlated with that of \cref{eq:res} and is accordingly propagated into the ``stat'' error using the $\Gamma$-method \cite{Wolff:2003sm,Ramos:2018vgu}.

The ratio in \cref{eq:res} can also be combined with $w_0^{\rm WP25}=0.17236\,{\rm fm}$, the value used to define the hadronic scheme adopted in the 2025 White Paper of the Muon $g-2$ Theory Initiative~\cite{Aliberti:2025beg}. Using our $N_f=2+1$ result at the Edinburgh physical point gives
\begin{equation}\label{eq:t0WP25}
	\left.\sqrt{t_0}\right|_{w_0^{\rm WP25}}
	=0.14244(25)_{\rm stat}(34)_{\rm cont}(14)_{\text{$\chi$-cont}}[44]_{\rm total}\;{\rm fm},
\end{equation}
where the quoted uncertainty comes from the conversion. Direct scale setting with $w_0$ has produced sizeable cutoff effects in some calculations of $a_\mu^{\rm HVP}$~\cite{Djukanovic:2024cmq,ExtendedTwistedMass:2026nbv}. The converted value offers a possible $\sqrt{t_0}$-based alternative, analogous to the $f_\pi$-based implementation in~\cite{ExtendedTwistedMass:2026nbv}; whether it reduces the final uncertainty remains to be tested.

\begin{figure}[t!]
  \centering
  \includegraphics[width=.45\textwidth]{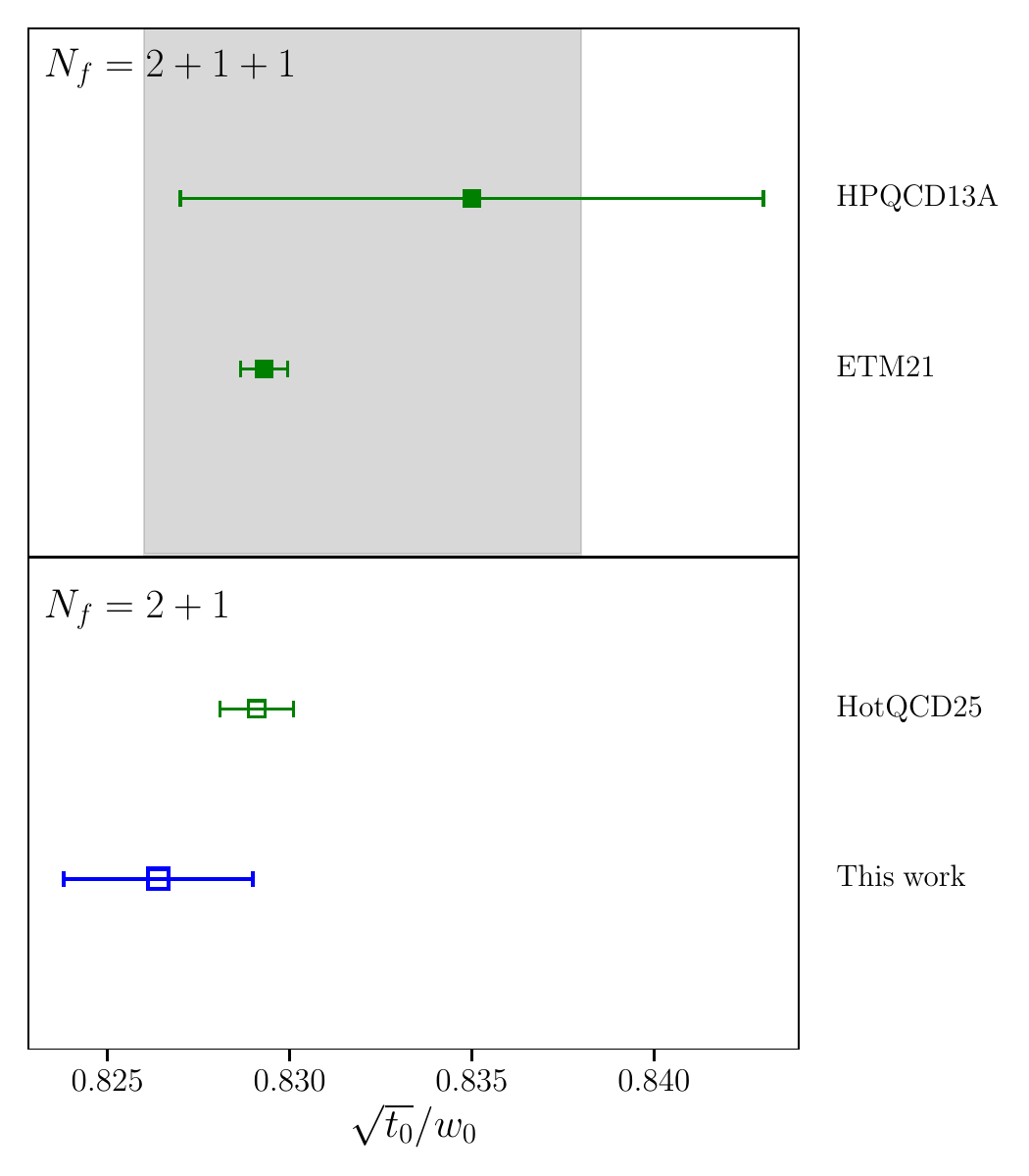}
  \includegraphics[width=.446\textwidth]{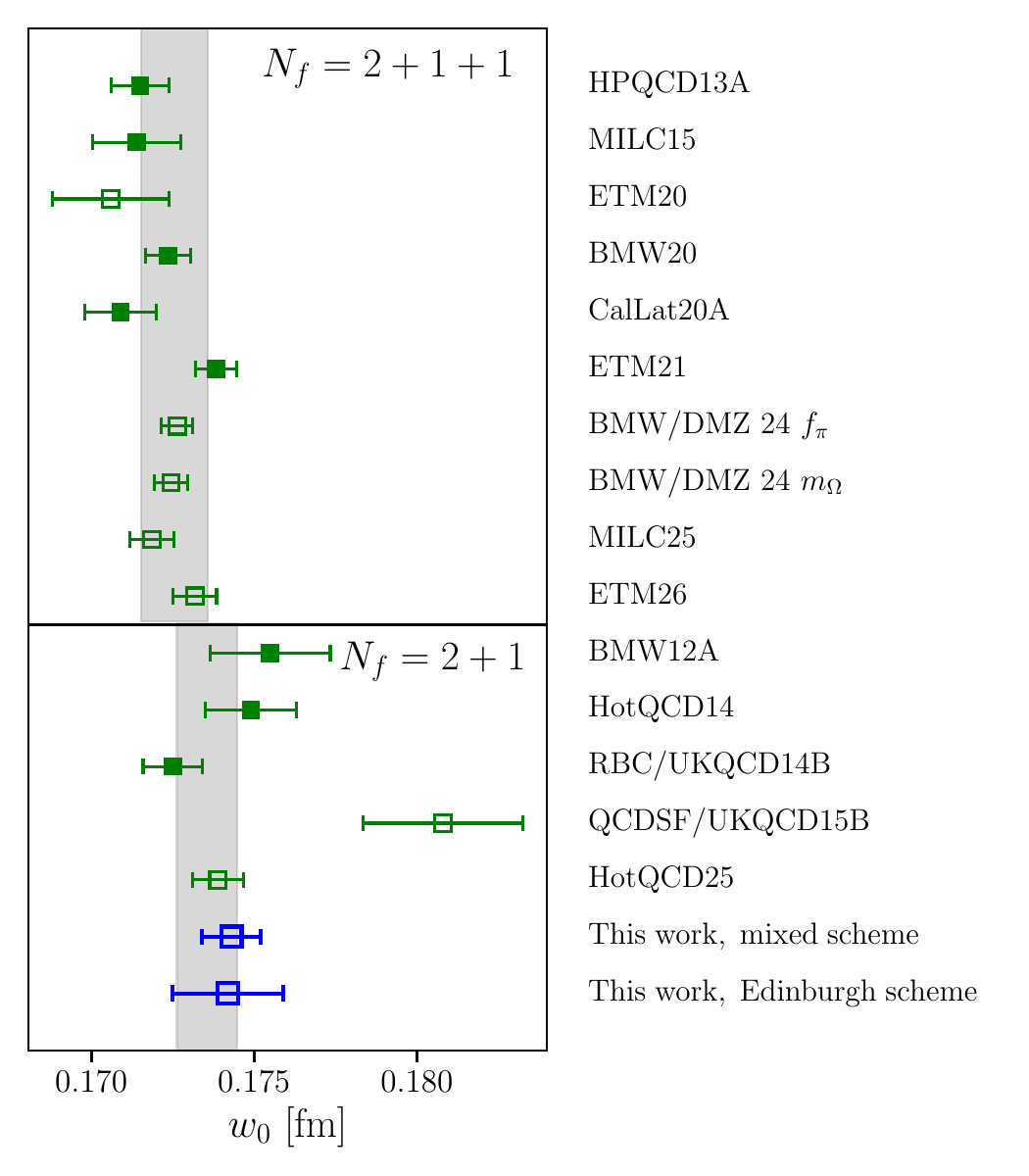}
  \caption{\label{fig:flag}
  Comparison of our results for $\sqrt{t_0}/w_0$ and $w_0$ compared to other determinations.
  The gray bands represent the FLAG averages \cite{FlavourLatticeAveragingGroupFLAG:2024oxs}.
  Upper side of plots is $N_f=2+1+1$ and lower side is $N_f=2+1$.
  ``BMW 12A'' refers to \cite{BMW:2012hcm}, ``HotQCD 14'' to \cite{HotQCD:2014kol}, ``RBC/UKQCD 14B'' to \cite{RBC:2014ntl}, ``QCDSF/UKQCD 15B'' to \cite{Bornyakov:2015eaa}, and ``HotQCD 25'' to \cite{Larsen:2025wvg}.
  ``HPQCD 13A'' refers to \cite{Dowdall:2013rya}, ``MILC 15'' to \cite{MILC:2015tqx}, ``ETM 20'' to \cite{ExtendedTwistedMass:2020tvp}, ``BMW 20'' to \cite{Borsanyi:2020mff}, ``CalLat 20A'' to \cite{Miller:2020evg}, ``ETM 21'' to \cite{ExtendedTwistedMass:2021qui}, ``BMW/DMZ 24'' to \cite{Boccaletti:2024guq}, ``MILC 25'' to \cite{FermilabLattice:2025dui} and ``ETM 26'' to \cite{ExtendedTwistedMass:2026nbv}.
  Filled symbols are results included in the FLAG average, while empty ones mark either points excluded from the latter or recent determinations not included in the latest FLAG review.
  }
\end{figure}

In \refig{fig:flag} we show our results compared to those of other works, together with the FLAG average \cite{FlavourLatticeAveragingGroupFLAG:2024oxs}.
Our value of $\sqrt{t_0}/w_0$ in \cref{eq:res} is consistent at the $1\sigma$ level with the determinations available in the literature.
The only other determination of this ratio in $2+1$ flavor QCD has been performed in \cite{Larsen:2025wvg} and is denoted by ``HotQCD 25'' in the figure.
It has been performed with HISQ fermion and Lüscher-Weisz gauge action on ensembles with unphysically large light quark masses, corresponding to $m_\pi\approx 160\,$MeV.
The authors estimate the effect of extrapolating to physical light quark masses using data at two lattice spacings and find it small at their present precision.
Their quoted continuum result nevertheless remains at $m_s/m_l=20$, without a corresponding shift or additional uncertainty.
Their result is compatible with ours.
The same is true for the two $2+1+1$ flavor results. 
The result denoted by ``HPQCD 13A'' \cite{Dowdall:2013rya} has been obtained using the HISQ action, whereas the precise result in ``ETM 21'' \cite{ExtendedTwistedMass:2021qui} has been obtained with Wilson-clover twisted-mass fermions and the Iwasaki gauge action at three lattice spacings down to $0.069\,$fm, using ensembles close to the physical quark masses.
Regarding $w_0^{\rm ph}$, we note that there is no consistent choice of scheme among the different results entering the comparison.
For $N_f=2+1$, our Edinburgh-consensus result in \cref{eq:w0_ph} agrees with the other determinations at the $1\sigma$ level, apart from QCDSF/UKQCD 15B \cite{Bornyakov:2015eaa}, which lies $2.7\sigma$ away.
The pattern is similar in the mixed scheme, where RBC/UKQCD 14B \cite{RBC:2014ntl} also differs at about $2\sigma$.
For $N_f=2+1+1$, our Edinburgh-consensus result is compatible with the other determinations at the $1$--$2\sigma$ level.
With the smaller uncertainty of the mixed-scheme result, several differences reach about $2$--$3\sigma$, while ETM 21 and ETM 26 remain compatible.
The central values of our two determinations of $w_0^{\rm ph}$ are nearly identical; the change in significance is driven by their different uncertainties.

The pattern of differences in $w_0^{\rm ph}$ broadly follows that
	observed among determinations of $\sqrt{t_0^{\rm ph}}$.

\section{Conclusions}
\label{sec:conclusions}

We have determined $\sqrt{t_0}/w_0$ in isospin-symmetric $2+1$-flavor QCD using 30 CLS ensembles at five lattice spacings between approximately 0.085\,fm and 0.039\,fm. 
The pion masses extend from about 420 MeV to the physical point. 
Wilson- and Zeuthen-flow data constrain the continuum extrapolation at the SU(3)-symmetric point; Wilson-flow data determine the remaining quark-mass dependence and residual cutoff effects.

In the physical-point scheme defined by the Edinburgh consensus, our result is
\begin{align}
  \label{eq:final}
  \left.\frac{\sqrt{t_0}}{w_0}\right|_{\rm ph}&=0.8264(15)_{\rm stat}(20)_{\rm cont}(8)_{\rm \chi-cont}[26]_{\rm total}.
\end{align}
and thus obtain a precision of about 0.31\%, which is limited by the contribution of systematic uncertainties from the continuum extrapolation.
In order to further reduce this uncertainty, the control over cutoff effects at the SU(3)-symmetric point has to be improved.
This could be achieved either by adding an additional, finer lattice spacing to the set of ensembles or by performing a combined continuum extrapolation of results obtained with multiple actions.
The latter would go beyond our approach of combining different discretizations of the flowed observables, as it would allow to vary the dynamical cutoff effects stemming from gauge and fermion action.
We note that it is not clear how much discretization effects for $\sqrt{t_0}/w_0$ differ between the different actions that are currently employed for large-scale simulations. 
Judging from the information provided in Figure 3 of \cite{Larsen:2025wvg} and Figure 11 of \cite{ExtendedTwistedMass:2021qui}, the approach to the continuum limit in these works looks quite similar to what is observed in this work, cf. Fig.~\ref{fig:continuum}.

\section*{Acknowledgements}

The authors want to thank Alberto Ramos, Gregorio Herdoiza and Carlos Pena for the very useful discussions and critical reading of a previous version of the manuscript.
We are grateful to our colleagues in the CLS initiative for sharing
ensembles.
The authors gratefully acknowledge the Gauss Centre for Supercomputing e.V. (\url{www.gauss-centre.eu}) for funding this project by providing computing time on the GCS Supercomputers HAWK at the
Höchstleistungsrechenzentrum Stuttgart (HLRS) and SuperMUC-NG at Leibniz Supercomputing Centre (\url{www.lrz.de}) to generate the gauge ensemble F300.
Further calculations for this project have been performed on the HPC
clusters HIMster-II at Helmholtz Institute Mainz and
Mogon-II at Johannes Gutenberg-Universität (JGU)
Mainz.
The authors gratefully acknowledge the computer resources at Artemisa and the technical support provided by the Instituto de Fisica Corpuscular, IFIC(CSIC-UV).
Artemisa is co-funded by the European Union through the 2014-2020 ERDF Operative Programme of Comunitat Valenciana, project IDIFEDER/2018/048.
This work is supported by the DFG through the Emmy Noether Programme (Project number 559290516), and by the Spanish Research Agency (Agencia Estatal de Investigación) through national project CNS2022-136005 and AEI/MCIU through grant PID2023-148162NB-C21 and ASFAE/2022/020

\cleardoublepage
\begin{appendices}
\section{Lattice observables}
\label{app:tables}

\begin{table}[tpbh!]
	\centering
	\begin{tabular}{l l l l l l l}
id & $am_{\pi}$ & $am_K$ & $af_{\pi}$ & $t_0/a^2$ & $w_0^2/a^2$ & $\sqrt{t_0}/w_0$ \\
\toprule
H101 & 0.18250(72) & 0.18259(72) & 0.06364(30) & 2.8472(55) & 3.6930(114) & 0.8780(6) \\
     &             &             &             & 2.5847(61) & 3.5479(127) & 0.8536(8) \\
H102 & 0.15383(81) & 0.19134(71) & 0.06044(30) & 2.8789(69) & 3.7560(136) & 0.8755(6) \\
H105 & 0.12155(115) & 0.20222(85) & 0.05781(84) & 2.8863(85) & 3.7757(166) & 0.8743(7) \\
N101 & 0.12120(56) & 0.20146(36) & 0.05773(35) & 2.8927(36) & 3.7924(78) & 0.8734(4) \\
C101 & 0.09570(79) & 0.20584(44) & 0.05511(35) & 2.9119(42) & 3.8332(92) & 0.8716(4) \\
C102 & 0.09640(87) & 0.21766(50) & 0.05507(46) & 2.8682(52) & 3.7562(112) & 0.8738(5) \\
D150 & 0.05654(94) & 0.20835(35) & 0.05216(31) & 2.9439(43) & 3.9034(98) & 0.8684(4) \\
\midrule
B450 & 0.16081(50) & 0.16101(50) & 0.05659(19) & 3.6628(133) & 4.8445(317) & 0.8695(11) \\
H400 & 0.16315(69) & 0.16335(69) & 0.05679(30) & 3.6339(131) & 4.7912(280) & 0.8709(11) \\
     &             &             &             & 3.3801(91) & 4.6195(215) & 0.8554(12) \\
S400 & 0.13503(46) & 0.17019(41) & 0.05399(34) & 3.6909(75) & 4.9040(167) & 0.8675(7) \\
N452 & 0.13546(30) & 0.17031(27) & 0.05440(14) & 3.6730(44) & 4.8615(110) & 0.8692(5) \\
N451 & 0.11064(46) & 0.17822(27) & 0.05205(15) & 3.6821(67) & 4.8964(137) & 0.8672(4) \\
D450 & 0.08346(52) & 0.18393(26) & 0.04969(15) & 3.6960(52) & 4.9326(116) & 0.8656(4) \\
D451 & 0.08338(35) & 0.19382(16) & 0.04978(21) & 3.6647(56) & 4.8789(138) & 0.8667(6) \\
D452 & 0.05932(60) & 0.18645(18) & 0.04749(15) & 3.7251(37) & 5.0056(96) & 0.8627(4) \\
\midrule
N202 & 0.13436(32) & 0.13438(32) & 0.04846(13) & 5.1524(195) & 6.9772(502) & 0.8593(13) \\
     &             &             &             & 4.9149(216) & 6.8306(556) & 0.8483(21) \\
N203 & 0.11249(27) & 0.14395(23) & 0.04643(16) & 5.1463(72) & 6.9814(206) & 0.8586(8) \\
N200 & 0.09221(29) & 0.15064(24) & 0.04420(18) & 5.1626(69) & 7.0373(176) & 0.8565(6) \\
D251 & 0.09200(14) & 0.15037(11) & 0.04448(09) & 5.1654(48) & 7.0388(129) & 0.8566(4) \\
D200 & 0.06502(28) & 0.15630(17) & 0.04237(20) & 5.1788(62) & 7.0982(172) & 0.8542(6) \\
D201 & 0.06498(43) & 0.16308(24) & 0.04263(25) & 5.1360(83) & 6.9946(212) & 0.8569(7) \\
E250 & 0.04232(22) & 0.15936(08) & 0.04018(11) & 5.2021(37) & 7.1756(104) & 0.8515(3) \\
\bottomrule
\end{tabular}
\caption{\label{tab:obs1}
Pion and kaon masses and pion decay constant, as computed in \cite{Djukanovic:2024cmq,Beltran:2026ofp}, together with the flow scales $t_0/a^2$ and $w_0^2/a^2$ -- using the Wilson flow -- and the ratio $\sqrt{t_0}/w_0$ for the set of ensembles entering the analysis.
A second line with additional values of the flow scales and their ratio is provided for ensembles for which the Zeuthen flow and improved discretization of the energy density is available.
Flow scales and their ratio are quoted before imposing the mass-shift to the LCP defined by $\phi_2^{\rm sym}=0.74$.
The pseudoscalar masses and decay constant have been corrected for finite size effects with the resummed finite-volume formulae of Refs.~\cite{Colangelo:2003hf,Colangelo:2005gd} in the case of pions, and at NLO in $\chi$PT for kaons.
Continued in \ret{tab:obs2}.
}
\end{table}

\begin{table}[tpb]
	\centering
	\begin{tabular}{l l l l l l l}
id & $am_{\pi}$ & $am_K$ & $af_{\pi}$ & $t_0/a^2$ & $w_0^2/a^2$ & $\sqrt{t_0}/w_0$ \\
\toprule
J307 & 0.10508(14) & 0.10510(14) & 0.03810(09) & 8.5883(120) & 11.9488(370) & 0.8478(8) \\
N300 & 0.10574(31) & 0.10588(31) & 0.03817(16) & 8.5547(312) & 11.8541(775) & 0.8495(15) \\
     &             &             &             & 8.3107(254) & 11.7629(529) & 0.8405(12) \\
J306 & 0.08685(12) & 0.11320(12) & 0.03643(08) & 8.5818(133) & 11.9474(411) & 0.8475(7) \\
J303 & 0.06467(22) & 0.11963(19) & 0.03439(16) & 8.6167(138) & 12.0882(527) & 0.8443(13) \\
J304 & 0.06561(23) & 0.13187(18) & 0.03418(14) & 8.4994(118) & 11.8092(326) & 0.8484(7) \\
E300 & 0.04399(12) & 0.12402(09) & 0.03264(12) & 8.6139(59) & 12.1180(249) & 0.8431(6) \\
F300 & 0.03403(12) & 0.12360(11) & 0.03195(13) & 8.6444(79) & 12.2182(339) & 0.8411(8) \\
\midrule
J500 & 0.08157(17) & 0.08165(17) & 0.02983(10) & 13.9649(390) & 19.7965(1336) & 0.8399(15) \\
     &             &             &             & 13.7285(288) & 19.6652(855) & 0.8355(12) \\
J501 & 0.06590(14) & 0.08794(13) & 0.02855(11) & 13.9802(511) & 19.8023(1198) & 0.8402(12) \\
\bottomrule
\end{tabular}
\caption{\label{tab:obs2}
Continued from Table~\ref{tab:obs1}.
}
\end{table}

\section{Fitting procedure}
\label{app:chisq}

Fits are performed by finding the parameters $\{p_i\}_{i=1}^{N_{\rm param}}$ that minimize the $\chi^2$ function
\begin{equation}
  \label{eq:chisq}
\chi^2=\sum_{i,j=1}^{N_{\rm dat}}\left(f(x_i; \boldsymbol{p})-y_i\right)W_{ij}\left(f(x_j; \boldsymbol{p})-y_j\right),
\end{equation}
where $f(x_i; \boldsymbol{p})$ is the fit function, $\{x_i\}_{i=1}^{N}$ the set of independent variables, $\{y_i\}_{i=1}^{N_{\rm dat}}$ the set of dependent variables, and $W_{ij}$ is the weight matrix.
We refer to the fit as correlated when $W=C^{-1}$, where $C$ is the covariance matrix of the $y$ data, and as uncorrelated when $W=\operatorname{diag}(C)^{-1}$.%
\footnote{Here, ``uncorrelated'' refers only to the diagonal weights in the $\chi^2$ fit. Correlations among observables and fit results from shared ensembles are retained in the statistical error propagation, including through any model average.}
Following \cite{Bruno:2022mfy}, one can compute the $\left<\chi^2\right>$ (the expected value of the $\chi^2$) and the p-value of a fit for an arbitrary $W$.
One can generalize the definition in \cref{eq:chisq} to the case in which the $x$-data also has errors by including additional auxiliary parameters $\{\tilde{p}_i\}_{i=1}^N$ as
\begin{equation}
  \label{eq:chisq_full}
  \chi^2=\sum_{i,j=1}^{N_{\mathrm{dat}}+N}\left(Y-F\right)_i \, \mathcal{W}_{ij}\, \left(Y-F\right)_j\,,
\end{equation}
where we defined the joint-data vector
\begin{equation}
  Y=(y_1,...,y_{n_{\mathrm{dat}}},x_1,...,x_{N}),
\end{equation}
and
\begin{equation}
	F_i =
	\begin{cases}
		f_i(\tilde p_1,\ldots,\tilde p_N;\boldsymbol p),
		& 1\leq i\leq N_{\rm dat},\\
		\tilde p_{i-N_{\rm dat}},
		& N_{\rm dat}<i\leq N_{\rm dat}+N .
	\end{cases}
\end{equation}
For a fully correlated fit, the weight matrix $\mathcal{W}$ in \cref{eq:chisq_full} must be the inverse of the full covariance matrix of the $Y$ vector, which we denote $\mathcal{C}$, containing the correlations of the $x$- and the $y$-data, and the correlations between the two.
In practice for our continuum extrapolation fits in Sec.~\ref{sec:continuum} and chiral-continuum extrapolation in Sec.~\ref{sec:chiral_extrap}, we perform an uncorrelated fit including the errors in the $x$-axis, i.e. we take the inverse of the diagonal of the $\mathcal{C}$ matrix.
Alternatively, we tested the impact of employing a correlated fit without $x$-errors, i.e. using \cref{eq:chisq} with $W=C^{-1}$, and found virtually the same main result as in \cref{eq:final}.

\end{appendices}

\cleardoublepage
\bibliographystyle{JHEP.bst}
\bibliography{biblio}

\end{document}